\documentclass[aps,pra,reprint,superscriptaddress,twocolumn]{revtex4-2}
\usepackage{adjustbox}

\usepackage[english]{babel}
\usepackage{ucs}
\usepackage[utf8x]{inputenc}
\usepackage{graphicx}
\usepackage{dcolumn}
\usepackage{bm}
\usepackage{bbm}
\usepackage{amsmath}
\usepackage{amssymb}
\usepackage{mathrsfs}
\usepackage{bm}
\usepackage{bbold}
\usepackage{empheq}
\usepackage{amsfonts}
\usepackage{color,colordvi}
\usepackage{physics}
\usepackage[colorlinks=true]{hyperref}
\usepackage[capitalize]{cleveref}
\usepackage{lipsum}
\usepackage{etoolbox}
\usepackage{comment}

\usepackage{xcolor}
\usepackage[normalem]{ulem}

\begin{document}
    \title{High-gain effects in broadband continuous-wave parametric down conversion sources and measurements with undetected photons}

    \author{Martin~Houde}
	\affiliation{Department of Engineering Physics, École polytechnique de Montréal, Montréal, QC, H3T 1J4, Canada}
    
    \author{Franz~Roeder}
	\affiliation{Paderborn University, Integrated Quantum Optics and Institute for Photonic Quantum Systems (PhoQS), 33095, Paderborn, Germany}
    
    \author{Christine~Silberhorn}
	\affiliation{Paderborn University, Integrated Quantum Optics and Institute for Photonic Quantum Systems (PhoQS), 33095, Paderborn, Germany}
    
    \author{Benjamin~Brecht}
	\affiliation{Paderborn University, Integrated Quantum Optics and Institute for Photonic Quantum Systems (PhoQS), 33095, Paderborn, Germany}

	\author{Nicolás~Quesada}
	\affiliation{Department of Engineering Physics, École polytechnique de Montréal, Montréal, QC, H3T 1J4, Canada}
	
	\begin{abstract}
        We study theoretically how high-gain effects affect the measurement outcome of visible signal spectra in undetected photon measurement schemes. We consider two interferometric configurations: firstly, the SU(1,1) interferometer where the idler incurs loss and additional dispersion in between two identical, lossless, squeezers; secondly,  the induced coherence interferometer where the idler incurs loss and additional dispersion in between two identical, lossless, squeezers and where the second squeezer is seeded by the idler and a vacuum ancilla mode. Furthermore, we consider a distributed loss configuration where the idler incurs loss as it propagates in the nonlinear medium. Motivated by experimental evidence and due to the fact that broadband sources are ideal for these measurement schemes, we use the dispersive data of a third-order dispersion engineered integrated waveguide parametric down conversion (PDC) source presented in \textit{New Journal of Physics 26, 123025 (2024)}\cite{Roeder2024ultrabroad} to model the PDC spectra in the three configurations. For each configuration we consider the case of idler-only (i) absorption, (ii) additional dispersion, and (iii) the combined effects. We obtain results which outline the strength and weaknesses of the different configurations at different operation points. 
	\end{abstract}
    \maketitle
	\section{Introduction}
	\label{sec:Intro}

    Undetected photon measurement schemes \cite{Lindner2021,Kaufmann2022,Neves2024,Paterova2018,Dong2025} have become an integral component of quantum spectroscopy. By leveraging nonlinear effects, one can infer absorptive or dispersive effects, imprinted by a sample to be probed, at wavelengths where detectors are inefficient or unavailable \cite{Lindner2021,Haase2023,Kutas2025}, such as in the mid-infrared (MIR) spectral range, by detecting signals efficiently in the optical or near-infrared parts of the spectrum.   

    The most common implementation for these measurements schemes is based on nonlinear interferometers such as the SU(1,1) \cite{Yurke1986} and the induced coherence (IC) \cite{Wang91} interferometers. These systems make use of broadband nonlinear sources that generate strongly frequency-time entangled pairs of nondegenerate signal and idler photons. By convention, the idler is generated in the undetectable range and probes a sample\textemdash called the analyte\textemdash of interest while the signal remains in a range that can be efficiently detected. For the SU(1,1) interferometer, the signal and idler photons are recombined at a second nonlinear crystal after the interaction with the sample has taken place. In this case, nonlinear interference imprints the information obtained by the idler onto the signal. For the IC interferometer, only the idler seeds the second nonlinear crystal after probing the sample. This in turn generates a new signal photon\textemdash called the ancilla photon\textemdash that carries information from the idler-probe interaction. One can then measure the ancilla or recombine the ancilla and signal modes via a beamsplitter to obtain the relevant information. While the IC configuration led to the first experimental realization of undetected measurement schemes \cite{Lemos2014} and has recently attracted renewed attention \cite{Gemmell2024,Roeder_OCT, volkoff2024radar}, SU(1,1) interferometers have become the central platform for broadband quantum spectroscopy \cite{Lindner2021,Panahiyan2022,Panahiyan2023,Neves2024,Kaufmann2022,Riazi2019dispersion, volkoff2025su11}.

    Conversely, an experimentally simpler albeit more restrictive distributed-loss (DL) scheme has recently garnered attention for undetected photon measurements\cite{Kumar2020,Krstic2023}. In this configuration, the idler interacts with the analyte, inducing loss and possibly modifying the dispersion, throughout the photon pair generation process. This directly imprints the information onto the signal spectrum, removing the need for interference. However, it requires analytes that can interact directly with the nonlinear medium. 

    Nonlinear waveguides are commonly used as the sources for the different sensing configurations to achieve broadband emission of nondegenerate photons. In these waveguides, it is common practice to employ dispersion engineering techniques \cite{Pollmann2024brightsource, roeder2024su11prx, Roeder2024ultrabroad, Houde2025lnbo} to modify the group velocities and group-velocity dispersions of the modes and in turn increase the bandwidth of the generated photons. 
    
    Although these sources have been demonstrated and studied in the low-gain regime, it is not clear how high-gain effects in conjunction with dispersion engineering modifies the spectra of the modes. Furthermore, it is also important to study if there is some possible interplay between the different sensing configuration and these dispersion engineered sources. Previous low- and high-gain studies of nonlinear interferometers \cite{Okamoto2020Loss,Sharapova2018,Scharwald2023,Hashimoto2024, houde2026quantum}  and distributed-loss absorption measurements \cite{Krstic2023} have been conducted. However, to our knowledge, a comprehensive quantitative comparison with a focus on the signal spectra signatures is still lacking for the three sensing configurations. 

    In this work, we investigate the response of the signal spectra to idler-only effects for the three sensing configurations when employing dispersion engineered sources. To render our results more tangible, we base them on the broadband waveguide source from Ref.\cite{Roeder2024ultrabroad}. Based on a periodically poled titanium in-diffused lithium-niobate waveguide, this source is optimized for generating photon pairs centered at $860\,\mathrm{nm}$ and $2750\,\mathrm{nm}$ with a spectral bandwidth of $23\,\mathrm{THz}$. Our theory framework allows us to recreate the spectrum generated in the source and to predict the source performance for mid-infrared spectroscopy with undetected photons. However, we emphasize that our results are general and readily applicable to any other source. We find that high-gain effects can shift and modify the observed oscillations in the SU(1,1) and IC configuration leading to an erroneous analysis of incurred additional dispersion of the idler (when using low-gain expressions of Ref.\cite{Riazi2019dispersion}). For the IC configuration, we find that an additional phase term in the oscillations can be detrimental for spectroscopic measurements of the recombined ancilla-signal mode (especially for highly engineered sources). However, in the high-gain limit the ancilla spectra, operating in the visible range, behaves to a good approximation like the idler spectra. In the DL configuration, we find interesting gain-dependent effects due to the distributed nature of the loss.
    
\begin{figure}
		\includegraphics[width=1\linewidth]{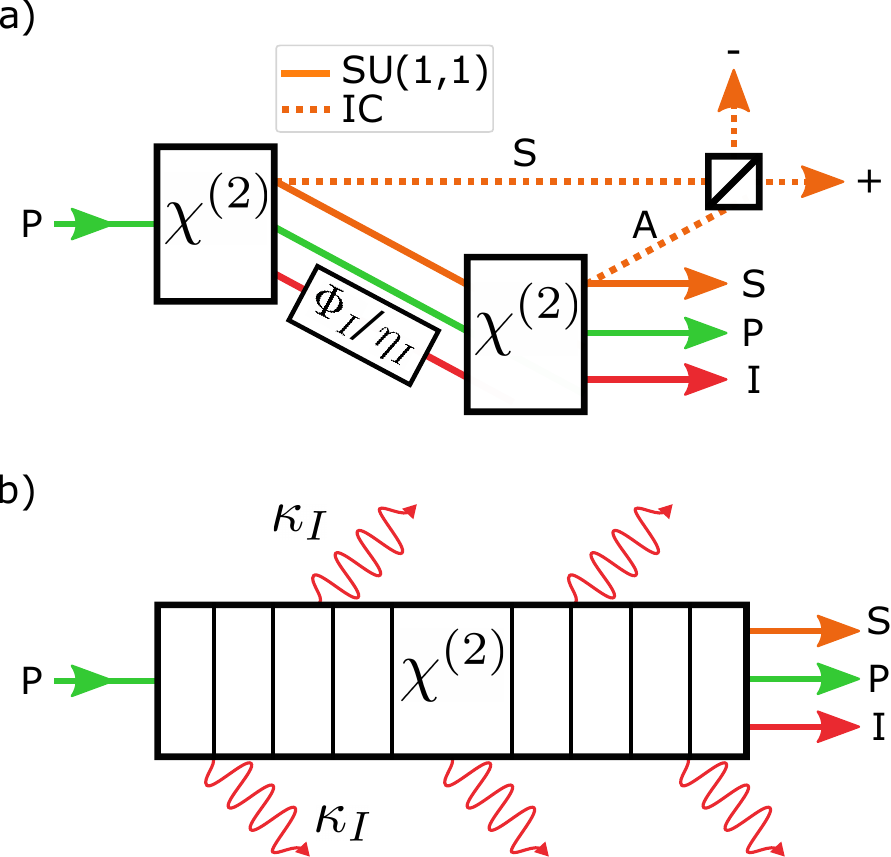}
		\caption{Schematic representation of the different sensing configurations and the respective loss implementations. (a) Interferometric sensing for SU(1,1) (solid orange lines) and IC (dotted orange line) models. Both nonlinear regions are taken to be identical and lossless. We treat losses via a beam splitter interaction (with transmission rate $\eta_{I}$ for the idler) between both nonlinear regions. To simulate optical path delays, we also allow the beam splitter interaction to induce additional dispersion for the idler mode ($\Phi_{I}$). SU(1,1) operation: both signal and idler seed the second nonlinear region. IC operation: only the idler seeds the second nonlinear region, newly generated ancilla mode is then combined with signal via a balanced beamsplitter. (b) DL model where the idler mode experiences decay rate ($\kappa_{I}$) as it propagates through the nonlinear region. Both losses and transmission rates are taken to be frequency dependent. P: pump; S: signal; I: idler; A: ancilla.}
		\label{fig:model}
	\end{figure}

    The remainder of this paper is organized as follows. In Sec.~\ref{sec:Model}, we motivate our study by summarizing experimental indications of intrinsic idler loss in nonlinear sources designed for interferometric operation. Section~\ref{sec:data} describes the simulation data used throughout the analysis. The treatment of idler-only loss is introduced in Sec.~\ref{sec:loss}. In Sec.~\ref{sec:cw}, we present the theoretical model and assumptions used for twin-beam generation in the continuous-wave limit. We consider the case of ideal lossless sources in Sec.~\ref{sec:lossless} and obtain analytic expressions for frequency-resolved second-order moments. Section~\ref{sec:intensity} discusses how dispersion characteristics influence the signal intensity distribution, reproducing features observed experimentally. Analytic results for the frequency-resolved second-order moments of the SU(1,1), IC, and DL schemes—depicted in Figs.~\ref{fig:model}(a) and (b)—are derived in Secs.~\ref{sec:su11}, \ref{sec:IC}, and \ref{sec:dist}, respectively. In Sec.~\ref{sec:absorption}, Sec.~\ref{sec:dispersion}, and Sec.~\ref{sec:interferometry} we study the signatures of idler-only loss, idler-only additional dispersion, and idler-only loss and additional dispersion, respectively, on the signal spectra for the three sensing configurations. Finally, we summarize our main findings in Sec.~\ref{sec:conc}.

	\section{Experimental motivation and model overview}
	\label{sec:Model}
    
    In Ref.\cite{Roeder2024ultrabroad}, some of the present authors demonstrated an ultra‑broadband photon‑pair source based on second‑order dispersion–engineered PDC in a periodically poled titanium in-diffused lithium‑niobate waveguide. Pumped by a continuous‑wave laser at $652~\,\mathrm{nm}$, the device generates signal and idler photons centered at $860~\,\mathrm{nm}$ and $2700~\,\mathrm{nm}$, respectively. Owing to group‑velocity matching and dispersion cancellation, it exhibits a spectral bandwidth of about $25~\,\mathrm{THz}$.
    
\begin{figure}
		\includegraphics[width=1\linewidth]{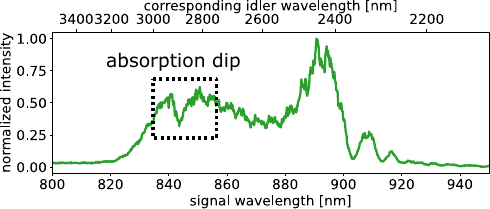}
		\caption{Absorption feature in the measured signal spectrum around $845\,\mathrm{nm}$, which corresponds to an idler wavelength of $2900\,\mathrm{nm}$. The measured dip in the intensity can be attributed to an absorption of the idler field in the waveguide material throughout the generation process.}
		\label{fig:absorption_spec}
	\end{figure}

    The measured signal spectrum [Fig.~\ref{fig:absorption_spec}] shows a pronounced asymmetry with a peak near $890~\,\mathrm{nm}$ and a dip at $845~\,\mathrm{nm}$, corresponding to an idler wavelength of $2900~\,\mathrm{nm}$. This dip is attributed to absorption by OH$^{-}$ impurities in lithium-niobate \cite{Kong1994} and can be interpreted as DL sensing, where frequency‑dependent idler attenuation during generation modifies the signal spectrum. This observation prompts the central question of whether an integrated quantum sensor based on undetected photons achieves optimal performance through nonlinear interferometry or through the DL mechanism itself when considering the signal spectra.

	\subsection{Data set and simulations}
	\label{sec:data}

    In this work, we use dispersion data generated for the source reported in Ref. \cite{Roeder2024ultrabroad}. These dispersion‑engineered waveguides are designed to minimize signal–idler group‑velocity mismatch and to cancel group‑velocity‑dispersion contributions in the phase‑matching function, requiring operation at a precise working point. A mode solver was used to obtain wavelength‑dependent Sellmeier equations for ordinary and extraordinary polarizations at a sample temperature of $T=230^{\circ}\mathrm{C}$. The continuous‑wave pump is set to $\bar{\lambda}_{P}=644~\,\mathrm{nm}$, with the pump and idler in ordinary and the signal in extraordinary polarization. Unless stated otherwise, the nonlinear interaction length is $L=40~\,\mathrm{mm}$, consistent with Ref.~\cite{Roeder2024ultrabroad}.
    
    We consider two poling periods that yield distinct signal spectra [Figs. \ref{fig:vacuum}(c,d)]. The ``flat parameters'', with $\Lambda_{\text{pol}} = 6.1879\,\mathrm{\mu m}$, represents an optimal operation point producing nearly uniform signal and idler spectra. The ``skewed parameters'',with $\Lambda_{\text{pol}} = 6.1883~\,\mathrm{\mu m}$, reproduces the measured spectrum from Fig.\ref{fig:absorption_spec}, corresponding to a slightly detuned, non‑optimal condition. In all subsequent analyses, we use the full dispersion relations derived from the generated data and the parameters specified above.
    
    Since the computed Sellmeier equations do not include material absorption, frequency‑dependent losses are introduced \textit{a posteriori}. Throughout the manuscript, the terms loss and absorption are used interchangeably to describe idler photon attenuation. For illustration, we implement two absorption peaks of different strength to highlight their respective influence on the signal spectra.

	\subsection{Loss implementation}
	\label{sec:loss}

    Here, we describe the implementation of spectrally dependent idler losses in our theory. For the DL model, we consider two absorption peaks of varying strengths at different wavelengths. As we are interested in cases where only the idler experiences loss, we set $\kappa_{S}(\omega)=0$---this corresponds to lossless propagation of the signal---and we take the idler decay rate to be of the form
    \begin{align}
    \kappa_{I}(\omega) =& \frac{-\log{(0.01)}}{L}e^{-(\omega-\omega_{1})/2\sigma_{1}^2}\nonumber\\
    &+\frac{-\log{(0.3)}}{L}e^{-(\omega-\omega_{2})/2\sigma_{2}^2}.
    \end{align}
    To obtain the same overall attenuation in the beam splitter model (SU(1,1) and IC configurations), we take 
    \begin{align}\label{eq:kappa_to_trans}
    \eta_{I}(\omega) = e^{-\kappa_{I}(\omega)L}
    \end{align}
    which corresponds to peak transmission rates of $1\%$ and $30\%$ at $\omega_1$ and $\omega_2$ respectively. This parametrization ensures that a change of interaction length in the DL model is mirrored in the SU(1,1) and IC configurations. We choose frequencies such that the idler absorption peaks are centered at $\lambda_{I,1}=2750~\,\mathrm{nm}$ with linewidth $\sigma_{\lambda,1}=20~\,\mathrm{nm}$ and at $\lambda_{I,2} = 2850~\,\mathrm{nm}$ with line-width $\sigma_{\lambda,2}=10~\,\mathrm{nm}$. When considering squeezing interactions, these idler wavelengths will correspond to signal wavelengths $\lambda_{S,1}\approx 841~\,\mathrm{nm}$ and $\lambda_{S,2}\approx 832~\,\mathrm{nm}$. All parameters are summarized in Tab.~\ref{tab:absorption_parameters} for convenience.

    \begin{table}[t]
        \centering
        \begin{tabular}{|c|c|c|c|}
            \hline
            Idler & Transmission & Bandwidth & Signal \\
            \hline
            $\lambda_{I,1} = 2750\,\mathrm{nm}$ & 1\% & $\sigma_{\lambda,1} = 20\,\mathrm{nm}$ & $\lambda_{S,1}\approx841\,\mathrm{nm}$ \\
            \hline
            $\lambda_{I,2} = 2850\,\mathrm{nm}$ & 30\% & $\sigma_{\lambda,2} = 10\,\mathrm{nm}$ & $\lambda_{S,2}\approx832\,\mathrm{nm}$ \\
            \hline
        \end{tabular}
        \caption{Summary of the parameters for the absorption peaks modelled throughout this manuscript.}
        \label{tab:absorption_parameters}
    \end{table}
    
	\section{Theory model}
	\label{sec:cw}

    In this section, we give a brief overview of the assumptions used to obtain analytic solutions to the second order moments generated by and throughout the different models. Furthermore, since the theory has been derived in detail in \cite{houde2026quantum}, we omit most of the derivation and state the analytic expressions for the relevant second order moments. 

    We assume that the nonlinear sources are identical and that they generate twin-beams via PDC. We assume that the sources are operated in a regime where self- and cross-phase modulations can be ignored. We label the modes $j\in{P,S,I,A}$ for pump, signal, idler, and ancilla, respectively. For each mode, we associate a central wavevector $\bar{k}_{j}$ with a central frequency $\bar{\omega}_{j}$ such that $\bar{k}_{j} = k_{j}(\bar{\omega}_{j})$ where $k_{j}(\omega)$ is the dispersion relation determined by the material and geometry of the squeezer. Note that since the sources are identical, the generated ancilla in the IC configuration has the same central frequency and wavevector as the signal. We consider Type-II PDC where we require that
    \begin{align}
		\bar{\omega}_{P}-\bar{\omega}_{S}-\bar{\omega}_{I}&=0,\label{Eq:EnconMT}\\
		\bar{k}_{P}-\bar{k}_{S}-\bar{k}_{I}&=0.\label{Eq:MomconMT}
	\end{align}
    For sources that are quasi-phase matched, the right-hand side of Eq.~(\ref{Eq:MomconMT}) should be changed to $\pm 2\pi/\Lambda_{\text{pol}}$ where $\Lambda_{\text{pol}}$ is the poling period. 
    
    We assume that the pump mode is prepared in a strong coherent state with a large number of photons which remains constant throughout the nonlinear interaction (undepleted classical pump approximation). In addition, we take the continuous-wave monochromatic (CW) pump limit where the pump spectral profile becomes proportional to a Dirac-delta. These assumptions considerably simplify the equations of motion allowing one to obtain analytic results. 

    As we are interested in how unmeasured idler-only effects act on the signal (or ancilla for IC), we focus on idler-only effects and assume that the signal (ancilla) are not affected by loss or additional dispersion. We begin by going over the interferometric configurations of Fig.~\ref{fig:model}(a) where the nonlinear regions are lossless.

    \subsection{Lossless continuous wave model}
    \label{sec:lossless}

     Under the assumptions presented above, the resulting ($z$, $\omega$) equations of motion for lossless twin-beam generation are\cite{houde2023sources,quesada2020theory,lipfert2018bloch,kolobov1999spatial,horoshoko2022generator}
    \begin{align}
     \frac{\partial}{\partial z}\hat{c}_{S}(\omega,z)&=i\Delta k_{S}(\omega) \hat{c}_{S}(\omega,z) + i\gamma \hat{c}^{\dagger}_{I}(-\omega,z),\label{eq:eom_cws}\\
     \frac{\partial}{\partial z}\hat{c}^{\dagger}_{I}(\omega,z)&=-i\Delta k_{I}(\omega) \hat{c}^{\dagger}_{I}(\omega,z) - i\gamma^{*} \hat{c}_{S}(-\omega,z)\label{eq:eom_cwi}.
    \end{align}
    where $\Delta k_{S/I}(\omega) = k_{S/I}(\omega) - \bar{k}_{S/I}$ are the dispersion relations detuned from their central wavevectors, $\gamma=|\gamma|e^{-i\Phi_{P}}$ represents the interaction strength and we have explicitly written out the dependence on phase of the pump. Note that the interaction parameter depends on many other material parameters~\cite{quesada2020theory,quesada2022BPP,horoshoko2022generator}. We use $x^{*}$ to denote the complex conjugate of $x$. In all the equations in this manuscript, the variable $\omega$ is used to indicate a detuning from a central frequency. Thus, for example, $\hat{c}_j(\pm\omega,z)$ denotes a bosonic Heisenberg evolved operator that destroys  photons of frequency $\bar{\omega}_j \pm \omega$. The equations of motion thus link the signal annihilation operator at frequency $\bar{\omega}_{S}+\omega$ to the idler creation operator at $\bar{\omega}_{I}-\omega$, a consequence of energy conservation, see Eq.~(\ref{Eq:EnconMT}).

    For numerical evaluation, we discretize frequency space such that $\omega_{n} = \omega_{0}+n\Delta\omega|^{N-1}_{0}$ for an $N$-size grid and introduce new, dimensionless operators $\hat{a}_{S/I}(\omega_{n}) = \hat{c}_{S/I}(\omega_{n})\sqrt{\Delta \omega}$. These new operators obey the same Heisenberg equations of motion (Eqs.(\ref{eq:eom_cws}) and (\ref{eq:eom_cwi}))

    We obtain analytic solutions to the equations of motions which allow us to relate Heisenberg evolved operators at the end of the nonlinear region to those before said region. From these operators we can construct the relevant second order moments by taking the expectation value with respect to (w.r.t.) vacuum, denoted $\langle \cdot \rangle_{\text{vac}} = \bra{\text{vac}}\cdot \ket{\text{vac}}$ with $\ket{\text{vac}}$ being the vacuum, of the appropriate operator combinations. Without loss of generality, we take the nonlinear region to extend from $z=0$ to $z=L$. When the signal and idler modes are initially in vacuum the only non-zero expectation values are $\langle\hat{a}_{\mu}(\omega,0)\left[\hat{a}_{\mu}(\omega,0)\right]^{\dagger} \rangle_{ \text{vac}}=1$ for $\mu\in{S,I}$ and the non-vanishing second order moments are 
     \begin{align}
         N^{\text{V}}(\omega) =& \langle\hat{a}^{\dagger}_{S}(\omega,L)\hat{a}_{S}(\omega,L) \rangle_{ \text{vac}}=\langle\hat{a}^{\dagger}_{I}(-\omega,L)\hat{a}_{I}(-\omega,L) \rangle_{ \text{vac}}\nonumber\\
         =&\left|\frac{2\gamma\sin{(\nu(\omega) L/2)}}{\nu(\omega)}\right|^{2}\label{eq:Nv},\\
         M^{\text{V}}(\omega) =& \langle\hat{a}_{S}(\omega,L)\hat{a}_{I}(\omega,L) \rangle_{ \text{vac}} \nonumber\\ =&2i|\gamma|e^{-i\Phi_{P}}\left[ \frac{\sin{(\nu(\omega) L/2)}}{\nu(\omega)}  \right]^{*}\biggl[ \cos{(\nu(\omega) L/2)} \nonumber\\
         &\left.+i\frac{\Sigma_{K}(\omega)}{\nu(\omega)}\sin{(\nu(\omega) L/2)}    \right]\label{eq:Mv_exp}.
     \end{align}
    where
    \begin{align}
        \Sigma_{K}(\omega) &=\Delta k_{S}(\omega) +\Delta k_{I}(-\omega),\\
        \nu(\omega) &= \sqrt{\left[\Sigma_{K}(\omega)\right]^2 -4\gamma^{2}}.\label{eq:nuK}
    \end{align}
    Note that $N^{\text{V}}(\omega)$ above represents the mean photon number on a band centered at frequency $\bar{\omega}_{S}+\omega$ with bandwidth $\Delta \omega$.
    
    With the expressions for the second order moments of a lossless source with vacuum inputs, we can construct the second order moments for the SU(1,1) and IC configurations. As we are considering twin-beams, the second order moments $N_{S}(\omega)$ and $N_{I}(-\omega)$ are equivalent, however, this is not the case when idler-only loss/additional dispersion is further implemented. For the rest of the manuscript we differentiate between both signal and idler moments and note that we also refer to these moments as \emph{intensities}.

	\subsubsection{Intensity distribution}
    \label{sec:intensity}

	\begin{figure}
		\includegraphics[width=1\linewidth]{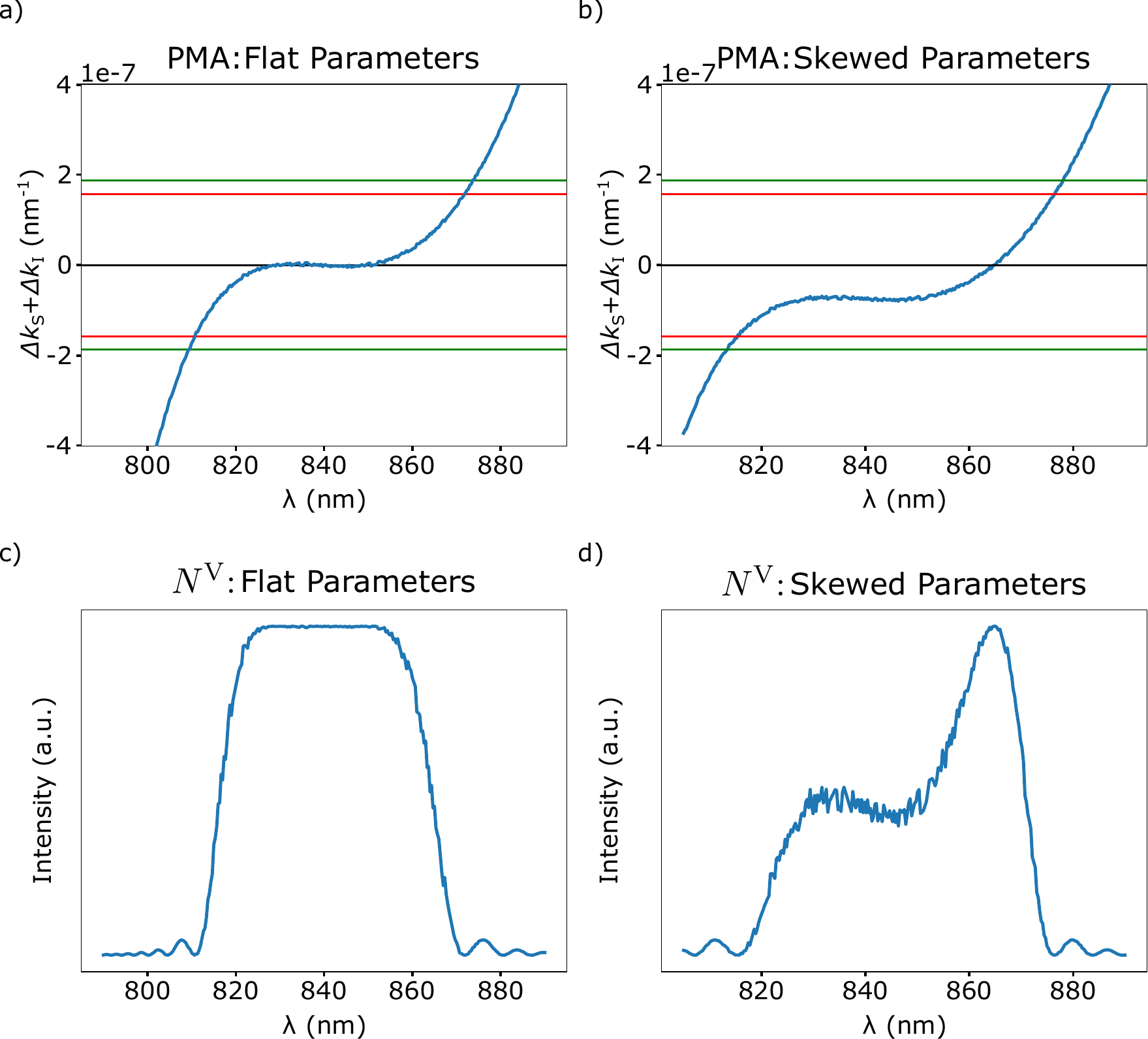}
		\caption{Phase-matching argument (PMA) and signal intensity curves for vacuum inputs for both sets of parameters. (a) PMA for the flat parameters. (b) PMA for the skewed parameters. We see that the PMA plateau is no longer phase-matched. (c) Signal intensity curve for vacuum inputs for the flat parameters. (d) Signal intensity curve for vacuum inputs for the skewed parameters. Although the PMA plateau is not phase-matched, we still see some amplification. The red(green) lines in (a) and (b) correspond to the PMA values of the first zero of the signal intensity curves for low(high)-gain regime.}
		\label{fig:vacuum}
	\end{figure}

    Before we start implementing loss, we look at how the signal intensity behaves for the two parameter sets mentioned in Sec.~\ref{sec:data} for vacuum inputs. The signal intensity is given by Eq.~(\ref{eq:Nv}). The dispersion data appears through the $\Sigma_{K}(\omega) =\Delta k_{S}(\omega) +\Delta k_{I}(-\omega)=k_{S}(\omega)+k_{I}(-\omega) - \bar{k}_{S} - \bar{k}_{I}=k_{S}(\omega)+k_{I}(-\omega) -\bar{k}_{P}$ term which is the usual term that we find inside the phase-matching function. We henceforth refer to $\Sigma_{K}(\omega)$ as the phase-matching argument (PMA).

    In Fig.~\ref{fig:vacuum}(a) and (b) we show the PMA for the flat and skewed parameters. For the flat set of parameters, we see that the PMA has a plateau region where we are well phase-matched. This gives rise to a broad and flat spectrum for the signal intensity as shown in Fig.~\ref{fig:vacuum}(c). Although the PMA curve is very similar for both cases, for the skewed set, the plateau in the PMA curve is de-tuned from zero and is therefore not phase-matched. However, when we consider the actual signal intensity, we see in Fig.~\ref{fig:vacuum}(d) that amplification/squeezing is still present. This gives rise to a skewed distribution where on the right we have a sharper peak structure, due to being phase-matched, and on the left we have a lower plateau, as is also seen in Fig.~\ref{fig:absorption_spec}. 
    
    It may seem surprising that we still have amplification/squeezing while not being phase-matched, however, when analyzing Eq.~(\ref{eq:Nv}) we find that this plateau remains within the region bounded by the first zeros of the signal intensity and is therefore not as heavily suppressed due the sinc-like structure of the signal intensity. In Figs.~\ref{fig:vacuum}(a) and (b) we include red(green) horizontal lines to show the values of the PMA for which $N^{\text{V}}(\omega)=0$ in the low(high)-gain. The relative height of the plateau region compared to the phase-matched peak depends on the relative distance from the PMA plateau to the $\Sigma_{K}(\omega)=0$ line. If we keep increasing the interaction strength, after a certain point we have that $|\text{PMA}|<2\gamma$ making $\nu(\omega)$ purely imaginary and giving rise to exponential growth for the plateau region. As long as the initially phase-matched peak does not grow too quickly before the plateau region starts to grow exponentially we observe signal intensity curves of the form of Fig.~\ref{fig:vacuum}(d).

    In the previous paragraph, we mentioned low(high)-gain regimes and throughout the manuscript we consider how gain affects certain results. We characterize the level of gain via the peak value of the signal intensity when no loss is present $N^{P}_{S} = \text{max}_{\omega}\left[N^{\text{V}}_{S}(\omega)\right]$ where ``P'' stands for peak. This peak value occurs when we have perfect phase-matching (i.e. $\Sigma_{K}(\omega)=0$) which gives $N^{P}_{S}=\sinh^{2}{(\gamma L)}$ and we tune the peak value by tuning the interaction strength $\gamma$. For the rest of the manuscript we choose $N^{P}_{S}=0.04$ for the low-gain regime and $N^{P}_{S}=14$ for the high-gain regime. Furthermore, to properly compare different intensities, we normalize all intensity curves such that the maximal value is unity.

	\subsection{SU(1,1) Model}
    \label{sec:su11}
	For the SU(1,1) model, we treat the sources as ideal squeezers without loss and allow for the idler mode to incur loss and additional dispersion in between these ideal squeezers. We treat the loss and additional dispersion via a beamsplitter interaction where the bath modes are taken to be Markovian quantum noise. We can then relate the second order moments after (``out'') the beamsplitter interaction to those before (``in'') as
    \begin{align}
        N^{\text{out,BS}}_{S}(\omega)&=N^{\text{in}}_{S}(\omega)\\
        N^{\text{out,BS}}_{I}(-\omega)&=\eta_{I}(-\omega)N^{\text{in}}_{I}(-\omega)\\
        M^{\text{out,BS}}(\omega)&=\sqrt{\eta_{I}(-\omega)}e^{i\Phi_{I}(-\omega) }M^{\text{in}}(\omega)
    \end{align}
    where $\eta_{I}(-\omega)$ is the frequency dependent idler transmission coefficient, $\Phi_{I}(-\omega)$ is the additional dispersion (from possible optical path delays (OPD) or external dispersive elements), and the superscript ``BS'' is to specify that we model a beamsplitter interaction. For the second pass, we use the nonlinear update rules for non-vacuum inputs of \cite{houde2026quantum} . 

    To construct the intensities for the full SU(1,1) configuration, we begin with a lossless source with vacuum inputs. We set $\Phi_{P}=0$ for the first pass setting it as a reference phase for the second pass. We implement the beamsplitter interaction and then feed the outputs into the second pass where we let $\Phi_{P}$ be arbitrary while keeping everything else identical. This arbitrary phase allows one to tune the functionality of the second pass (e.g. anti-squeezing operation) independently of the additional idler phases. We find that the signal and idler intensities at the end of the SU(1,1) interferometer are
    \begin{widetext}
    \begin{align}
        N^{\text{SU(1,1)}}_{S}(\omega)=&2N^{\text{V}}(\omega)+\left[N^{\text{V}}(\omega)\right]^{2}(1+\eta_{I}(-\omega)) \nonumber\\&-2\sqrt{\eta_{I}(-\omega)}\cos\left( \Phi(\omega) +\Psi(\omega)  \right)\left| M^{\text{V}}(\omega) \right|^{2},\label{eq:ns_su11}\\
        N^{\text{SU(1,1)}}_{I}(-\omega)=&N^{\text{V}}(\omega)(1+\eta_{I}(-\omega))+\left[N^{\text{V}}(\omega)\right]^{2}(1+\eta_{I}(-\omega)) \nonumber\\&-2\sqrt{\eta_{I}(-\omega)}\cos\left( \Phi(\omega) +\Psi(\omega)  \right)\left| M^{\text{V}}(\omega) \right|^{2}.   
    \end{align}
    \end{widetext}
    where we have grouped the idler and pump phases as $\Phi(\omega)=\Phi_{I}(-\omega)-\Phi_{P}$ and the additional phase
    \begin{align}\label{eq:psi_full}
        \Psi(\omega) = 2\tan^{-1}\left[\frac{\Sigma_{K}(\omega)}{\nu(\omega)}\tan{\left[\nu(\omega)L/2 \right]}     \right] +\pi 
    \end{align}
    comes from $\left[M^{\text{V}}(\omega)\right]^2$. These solutions will allow us to study the signal spectrum at the output of the SU(1,1) interferometer for different analytes and operation conditions. 
    
    Without loss and additional dispersion, we can set the second pass to act as a perfect anti-squeezer when we are phase-matched (i.e. $\Sigma_{K}(\omega)=0$) by setting $\Phi_{P}=\pm \pi$. In this case, the phase term $\Phi(\omega) +\Psi(\omega) = \pi\pm\pi$ and using the fact that $|M^{V}(\omega)|^{2} = N^{V}(\omega)\left( N^{V}(\omega)+1 \right)$ (see Eq.~(\ref{eq:Mv_exp})) we have that $N^{\text{SU(1,1)}}_{S}(\omega)=N^{\text{SU(1,1)}}_{I}(\omega)=0$. Note that if the system is not phase-matched this is no longer the case as $\Psi(\omega)\neq 0$ and one needs to be more careful to reach the anti-squeezing operation point. When loss is included we can no longer make the intensities vanish and we thus expect noticeable signatures on the signal intensity at the corresponding frequencies that idler loss is present. Similarly, we can set the second pass to act as an ideal squeezer by setting $\Phi_{P}=0$ when we are phase-matched without loss and additional dispersion. This in turn makes the system equivalent to a single pass of length $2L$ (assuming vacuum in between the passes).

	\subsection{Induced Coherence model}
	\label{sec:IC}
    Similarly to the SU(1,1) configuration, we treat the two sources as lossless, with the same pump-phase convention, and we implement the idler-only effects via a beamsplitter interaction for the IC model. However, unlike the SU(1,1) configuration, only the idler mode seeds the second pass and it interacts with a third vacuum mode, which we call the ``ancilla'' mode labelled with subscript ``A''. 

    Using the nonlinear and beamsplitter update rules mentioned above, we can construct solutions to the intensities after the second pass of the IC configuration. We find that the intensities are \cite{Kolobov2017IC, houde2026quantum}
    \begin{align}
        N^{\text{IC}}_{S}(\omega) &= N^{\text{V}}(\omega),\\
        N^{\text{IC}}_{I}(-\omega) & =N^{\text{V}}(\omega)+\eta_{I}(-\omega)N^{\text{V}}(\omega)(1+N^{\text{V}}(\omega)),\label{eq:ic_idler}\\
        N^{\text{IC}}_{A} (\omega)&= N^{\text{V}}(\omega)(1+\eta_{I}(-\omega)N^{\text{V}}(\omega)).\label{eq:ancilla}
    \end{align}
    Due to the induced coherence, we also find a non-zero mixed second order moment of the form 
    \begin{align}
        N^{\text{IC}}_{S,A}(\omega) =& \langle \hat{a}^{\dagger}_{S}(\omega;L)\hat{a}_{A}(\omega;L)  \rangle_{\text{vac}}\nonumber\\
         =& \sqrt{\eta_{I}(-\omega)}\nonumber\\ & \cdot N^{\text{V}}(\omega)\sqrt{1+N^{\text{V}}(\omega)}e^{i(\Delta K(\omega)L/2-\Phi(\omega)-\Psi(\omega)/2)}, 
    \end{align}
    where the new phase term, $\Delta K(\omega) = \Delta k_{S}(\omega) -\Delta k_{I}(-\omega)$, arises due to the fact that the signal does not go through the second pass. 

    By construction of the IC configuration, the ancilla operates at the same frequencies as the signal and they can therefore be recombined. In typical experimental measurement schemes one recombines the signal and ancilla modes via a 50:50 balanced beamsplitter and analyzes the intensity of one of the resulting beamsplitter arms. 
    Labelling the modes at the output of the balanced beamsplitter with subscripts ``$\pm$'', we find that the intensities are
    \begin{align}
        N^{\text{IC}, \text{BBS}}_{\pm}(\omega) =& \frac{1}{2}\left[ N^{\text{IC}}_{S}(\omega)+N^{\text{IC}}_{A}(\omega) \pm 2\text{Im}\left[ N^{\text{IC}}_{S,A} (\omega) \right]   \right] \nonumber\\
        =&\frac{1}{2}\biggl[2N^{V}(\omega)+\eta_{I}(-\omega)\left[N^{V}(\omega)\right]^{2}    \nonumber\\ &\pm 2\sqrt{\eta_{I}(-\omega)}N^{V}(\omega)\sqrt{1+N^{V}(\omega)}\nonumber\\ &\hspace{4mm}\cdot\sin(\frac{\Delta K(\omega) L}{2}-\Phi(\omega)-\Psi(\omega)/2)\biggr]\label{eq:IC_bbs}
    \end{align}
    where the superscript ``BBS'' is to specify that we have recombined the signal and ancilla modes.

	   \begin{figure}[ht]
	 	     \includegraphics[width=1\linewidth]{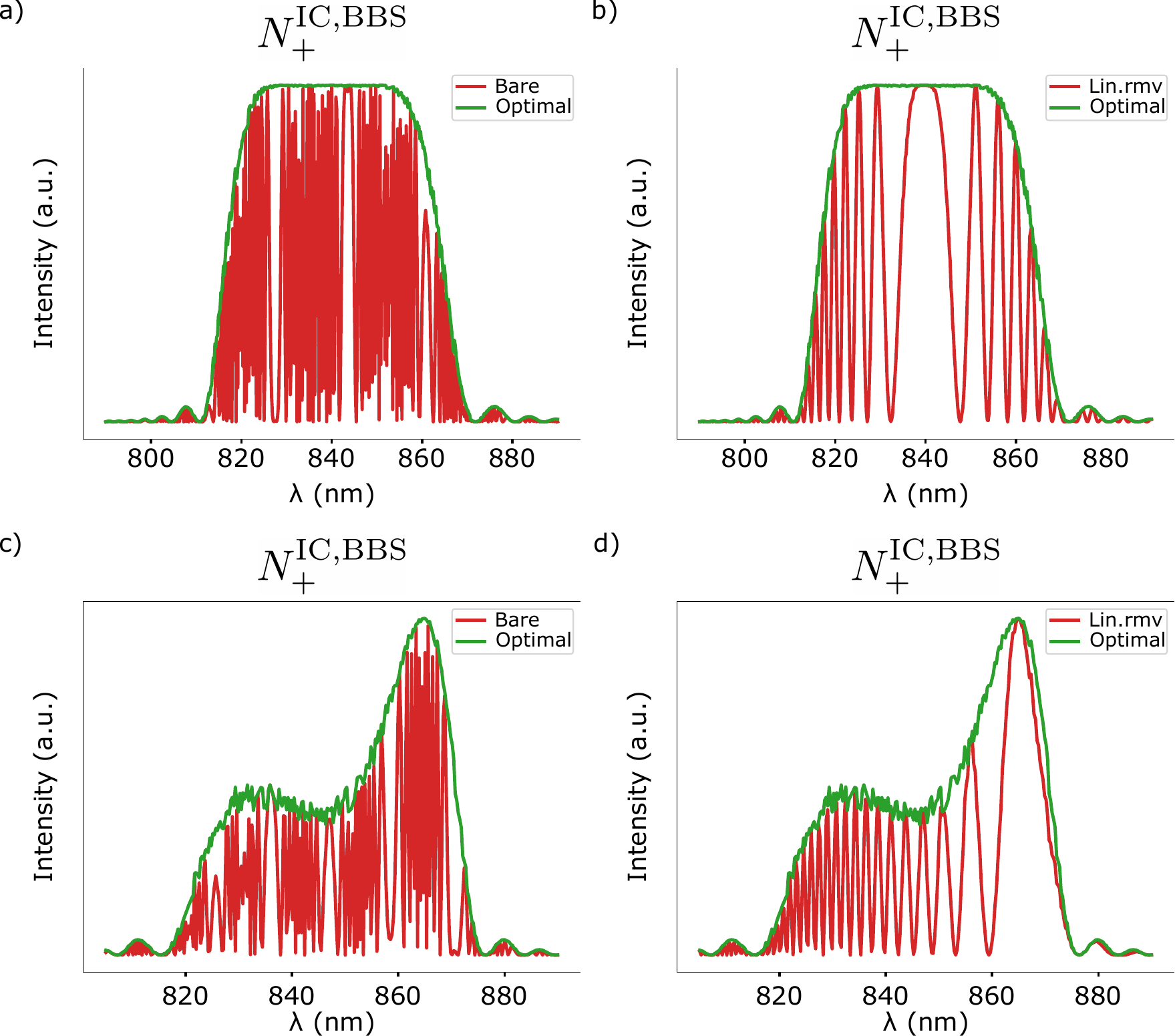}
	 	     \caption{Normalized intensity for the beamsplitter arm ($N^{\text{IC,BBS}}_{+}$) of the IC interferometer. Flat parameter set (a) without additional idler phases and (b) where additional idler phases cancel the linear contribution of $\Delta K(\omega)L/2$ (c.f. Sec.~\ref{sec:IC}). Skewed parameter set (c) without additional idler phases and (d) where additional idler phases cancel the linear contribution of $\Delta K(\omega)L/2$. Green optimal curves represent the situation where the $\Delta K(\omega)L/2$ term is completely negated.}
	 	\label{fig:ic_oscillation}
	   \end{figure}

    There are key differences between the IC beamsplitter arm intensity and the SU(1,1) signal intensity. First, the oscillations in the IC configuration are accompanied by an additional phase term $\Delta K(\omega)L/2$ which is only zero at the central frequencies. Depending on the sources, this could lead to very fast oscillations which may complicate the measurement operation. Indeed, in Figs.~\ref{fig:ic_oscillation}(a) and (c) we show the $N^{\text{IC,BBS}}_{+}(\omega)$ intensity for the flat and skewed parameters, respectively, and the oscillations are extremely fast. The green envelope-curve represents a theoretical optimal case where additional phases $\Phi(\omega)$ are chosen to completely cancel the $\Delta K(\omega)L/2$ term. Secondly, the oscillatory term comes with a different dependence on the single pass intensity $N^{V}(\omega)$. This in turn makes it very challenging for the intensity to vanish (i.e. have the second pass act as a perfect anti-squeezer). Indeed, for the lossless case, we find that the beamsplitter arm intensity only vanishes in the low-gain limit when $N^{V}(\omega)\ll 1$ such that $\sqrt{1+N^{V}(\omega)}\approx 1+N^{V}(\omega)/2$.
    
    For the ultra-broadband source considered in this manuscript we have that the group velocities are $v\equiv v_{S}=v_{I}$ and group-velocity-dispersions are $\beta \equiv \beta_{S}=-\beta_{I}$ which leads to 
    \begin{align}
        \Delta K(\omega)&\approx \left(\frac{1}{v_{S}}+\frac{1}{v_{I}}  \right)\omega+\frac{1}{2}\left( \beta_{S}-\beta_{I} \right)\omega^2\nonumber\\
        &\approx \frac{2}{v}\omega + \beta\omega^2\approx 2\Delta k_{S}(\omega).
    \end{align}
    This term is effectively proportional to twice the signal dispersion and induces very fast oscillations (as seen in Figs.~\ref{fig:ic_oscillation}(a) and (c)). By properly tuning the free-space propagation lengths of the different modes, one could imagine using $\Phi(\omega)$ to at least cancel out the linear contribution of $\Delta K(\omega)$. In Figs.~\ref{fig:ic_oscillation}(b) and (d) we remove the linear contribution of $\Delta K(\omega)L/2$. The quadratic term still leads to noticeable, albeit reduced, oscillations. This is specific to the source used in this manuscript and caused by the dispersion engineering. For less engineered sources, the group-velocity-dispersion terms could cancel out or give rise to a weaker quadratic term. Depending on the operation bandwidth, these weaker terms could be negligible (this is the case for the source used in Ref.~\cite{Roeder_OCT}). Therefore, although the IC configuration leads to additional oscillations, one can use the free-space propagation phases to cancel out the linear contribution of these oscillations. For experimental purposes it may be easier to add OPDs to the signal and the ancilla mode to cancel said linear term. Although we have focused the derivation with additional idler dispersion in mind, additional signal/ancilla dispersion would also appear in the $\Phi(\omega)$ term (similarly for the SU(1,1) interferometer). The effects of the quadratic and higher order terms need to be characterized on a case by case basis. Furthermore, one could also consider adding additional dispersive elements in the free-space propagation to cancel out the quadratic contributions, however, this would require significant calibration.

	\subsection{Distributed loss model}
	\label{sec:dist}
	For the DL model, we follow the derivations of Ref.~\cite{Caves1987quantum} and the similar models used in~\cite{houde2019loss,kopylov2025loss} to obtain spatial Heisenberg-Langevin equations of motion. For the same underlying assumptions as mentioned previously, the equations of motion are
    \begin{align}
        \frac{\partial}{\partial z}\hat{a}_{S}(\omega,z)=& i\Delta k_{S}(\omega) \hat{a}_{S}(\omega,z)+ i\gamma \hat{a}^{\dagger}_{I}(-\omega,z)\label{eq:eom_s_d}\\
        \frac{\partial}{\partial z}\hat{a}^{\dagger}_{I}(-\omega,z)=&-i(\Delta k_{I}(-\omega)-i\kappa_{I}(-\omega)/2 )\hat{a}^{\dagger}_{I}(-\omega,z) \nonumber\\& - i\gamma \hat{a}_{S}(\omega,z)+\sqrt{\kappa_{I}(-\omega)}\hat{b}^{\dagger}_{I}(-\omega,z)\label{eq:eom_i_d},
    \end{align}
    where we now have an idler decay rate $\kappa_{I}(\omega)$, with units of inverse length, and its associated bath mode $b_{I}(\omega,z)$ which we take to be Markovian quantum noise (i.e. $\langle b_{I}(\omega) \rangle=0$ and $\langle b_{I}(\omega,z)b^{\dagger}_{I}(\omega',z') \rangle=\delta(\omega-\omega')\delta(z-z')$). The pump-phase is set to zero. Note that apart from the additional noise term, the DL equations of motion above are very similar to the lossless case where instead we have $K_{I}(-\omega)=\Delta k_{I}(-\omega) -i\kappa_{I}(-\omega)/2$.
    
    We solve the equations of motion by employing a Green's-function-method formalism as done in Ref.~\cite{Caves1987quantum} and we obtain expressions for the signal and idler intensities. For a region of length $L$ with vacuum inputs we find
    \begin{align}
        N^{\text{DL}}_{S}(\omega)=&N^{\text{DL,V}}(\omega;L)\nonumber\\&+\kappa_{I}(-\omega)\int_{0}^{L}dz'N^{\text{DL,V}}(\omega;L-z'),\label{eq:distributed_signal} \\
        N^{\text{DL}}_{I}(-\omega)=&N^{\text{DL,V}}(\omega;L)\label{eq:distributed_idler}
    \end{align}
    where we have that
    \begin{align}
        N^{\text{DL,V}}(\omega;z)=& e^{-\kappa_{I}(-\omega)z/2}\left|\frac{2\gamma\sin{(\tilde{\nu}(\omega) z/2)}}{\tilde{\nu}(\omega)}\right|^{2} ,\label{eq:Nsvdist} 
    \end{align}
    and we have defined
    \begin{align}
        \tilde{\Sigma}_{K}(\omega) &= \Delta k_{S}(\omega)+\Delta k_{I}(-\omega)-i\kappa_{I}(-\omega)/2,\\
        \tilde{\nu}(\omega) &= \sqrt{\left[\tilde{\Sigma}_{K}(\omega)\right]^{2}-4\gamma^{2}} \label{eq:lossmatching}.
    \end{align}
    Note that if we set $\kappa_{I}=0$ in Eqs.~(\ref{eq:distributed_signal}) and (\ref{eq:distributed_idler}), we regain the solutions to the lossless case (Eqs.~(\ref{eq:Nv}) and (\ref{eq:Mv_exp})) as one would expect. When we are phase-matched, we have that $\tilde{\Sigma}_{K}(\omega)=-i\kappa_{I}(-\omega)/2$ which in turn makes $\tilde{\nu}(\omega) =i\sqrt{4\gamma^{2}+(\kappa_{I}(-\omega)/2)^{2}}$. Interestingly, this tells us that distributed loss enhances the interaction strength when we are phase-matched (however, the moments are also accompanied by an exponentially suppressing term).
    
    For the rest of the manuscript, we refer to the integral contribution in Eq.~(\ref{eq:distributed_signal}) as the ``added noise'' contribution (the other term is referred to as the ``bare'' contribution). The idler intensity has no added noise contribution, since the signal mode has no distributed loss, and the idler loss is present through the exponential decay term and the modified interaction parameter $\tilde{\nu}(\omega)$.

    \section{Absorption effects}\label{sec:absorption}
    With the expressions for the various intensities at hand, we now study how idler-only effects affect the intensities for the different configurations. In this section, we focus solely on the effects of idler loss for different levels of gain. We do not consider any additional phases and as such set $\Phi_{I}(-\omega)=\Phi_{P}=0$. 
    
    \subsection{SU(1,1) with idler absorption}\label{sec:su11_lost}
\begin{figure}[ht]
	 	     \includegraphics[width=1\linewidth]{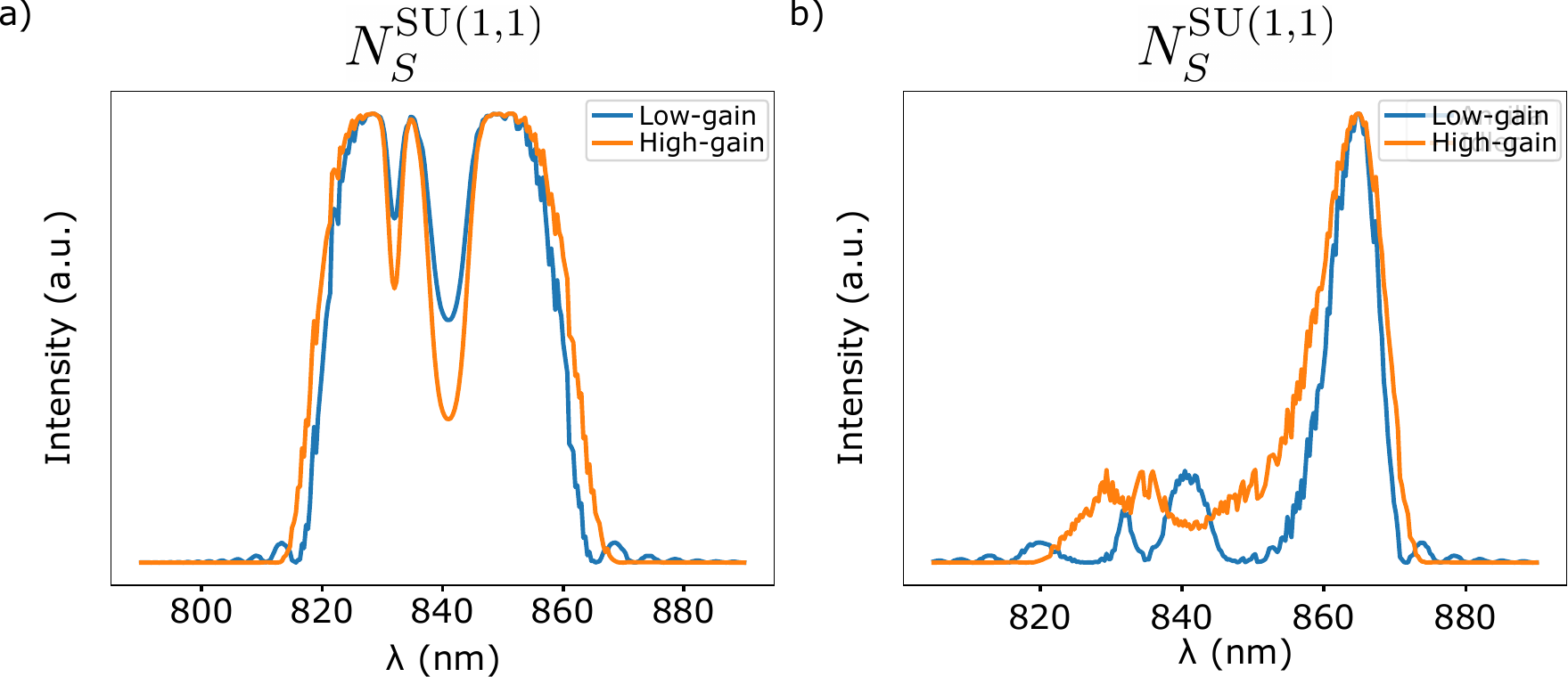}
	 	     \caption{Normalized signal intensity after an SU(1,1) inteferometer for the (a) flat and (b) skewed parameter sets in the low- and high-gain regimes. Near phase-matching in the flat parameter set gives rise to the second pass acting as a near-perfect squeezer and enhances the idler loss signature on the signal intensity. Lack of phase-matching in the skewed parameter set gives rise to gain-dependent operation of the second pass: in the low-gain regime, the second pass acts as an anti-squeezer in the plateau region.}
	 	\label{fig:su11_intensity}
	   \end{figure}

    We now look at the effects of idler-only loss for the SU(1,1) configuration. In Fig.~\ref{fig:su11_intensity} we show the normalized signal intensity at the end of the SU(1,1) interferometer (see Eq.~(\ref{eq:ns_su11})) for the (a) flat parameter set and the (b) skewed parameter set in both the low- and high-gain regime. For the flat parameter set, since we are phase-matched, the second pass acts approximately as a perfect squeezer. As expected from Eq.(\ref{eq:ns_su11}), the signal intensity is modified by the presence of idler loss. Furthermore, we see that in the high-gain regime the effects of idler absorption are more pronounced. This can be attributed to the fact that the $\left[ N^{\text{V}}(\omega)\right]^2$ term becomes dominant and it is tied to the idler transmission rate.
    
    For the skewed parameter set, since we are not phase-matched in the plateau region, the second pass does not act as a perfect squeezer and we get some gain-dependent interferometric effects. The effects are caused by the additional phase $\Psi(\omega)$. In the low-gain, the absorption effects are quite noticeable. In fact, in the low-gain, the second pass acts as an anti-squeezer in the plateau region. This gives rise to peaks in the intensity where the absorption occurs. In the high-gain, this is no longer the case and we see diminished signatures of the idler absorption. The actions of the second squeezer will be discussed in greater detail when we discuss interferograms in Sec.~\ref{sec:interferometry}. Note that these results are specific to the dispersions described in Sec.~\ref{sec:data}.

    \subsection{IC with idler absorption}\label{sec:ic_lost}
    For the IC configuration, we consider the effects of idler-only loss on both the ancilla and one of the beamsplitter arms.

    \subsubsection{Ancilla intensity}\label{sec:ancilla}
\begin{figure}[ht]
	 	     \includegraphics[width=1\linewidth]{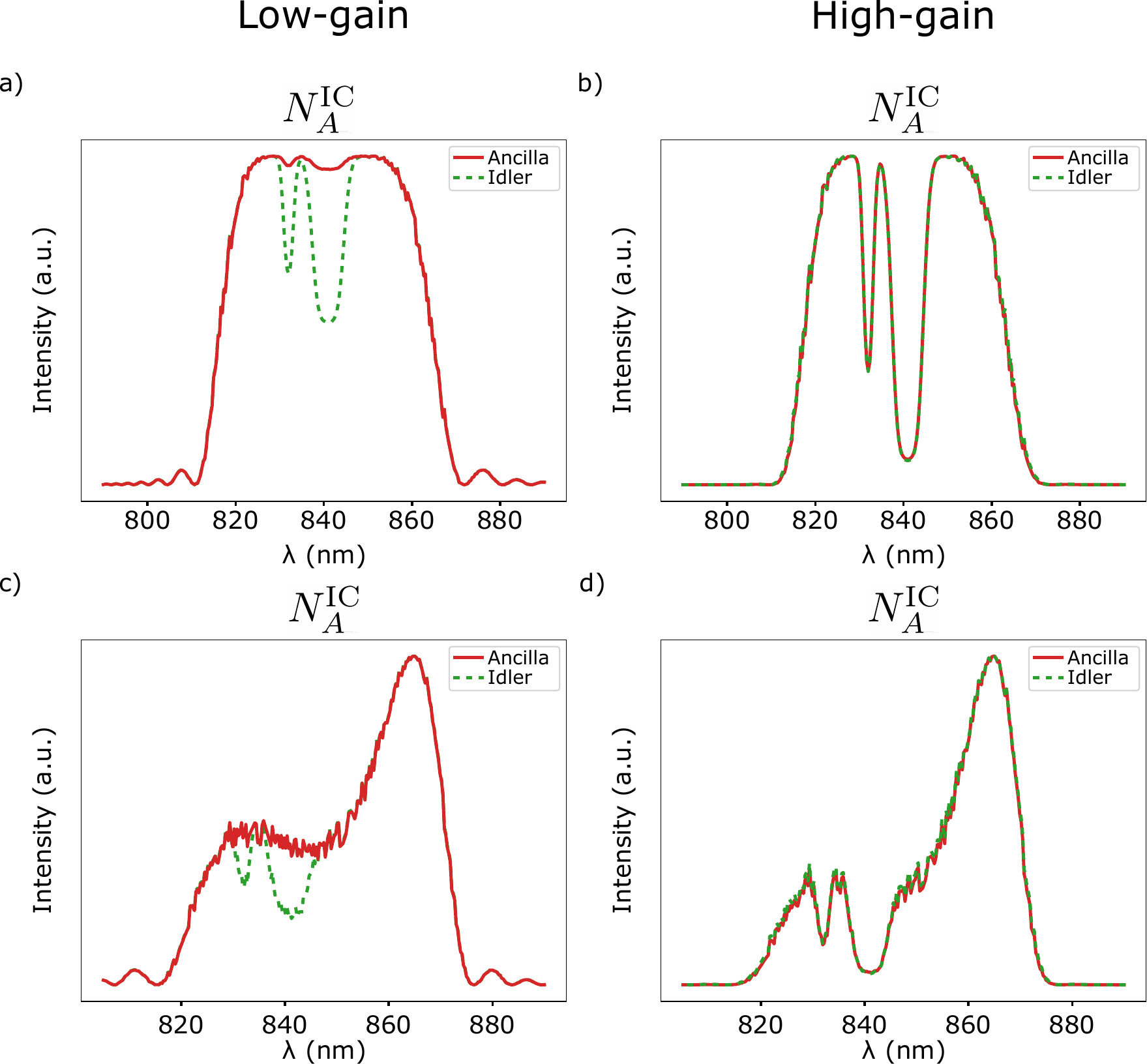}
	 	     \caption{Normalized ancilla intensity after the second pass of the IC interferometer. Flat parameter set in the (a) low- and (b) high-gain regimes. Skewed parameter set in the (c) low- and (d) high-gain regimes. For each plot we superpose the normalized idler intensity after the second pass. In the high-gain regime, the normalized ancilla intensity closely resembles the normalized idler intensity.}
	 	\label{fig:ancilla}
	   \end{figure}

    In Fig.~\ref{fig:ancilla}, we show the normalized ancilla intensity (see Eq.~(\ref{eq:ancilla})) for the flat parameter set in the (a) low- and (b) high-gain regimes and the skewed parameter set in the (c) low- and (d) high-gain regimes. For each plot, we superpose the corresponding normalized idler intensity (see Eq.~(\ref{eq:ic_idler})) at the end of the IC interferometer (dashed green line). 
    
    In the low-gain regime, we see that the idler loss effects on the ancilla intensity are quite small for the flat parameter set (Fig.~\ref{fig:ancilla}(a)) and unnoticeable for the skewed parameter set(Fig.~\ref{fig:ancilla}(c)). This can be explained by the fact that the idler transmission rate is proportional to $\left[ N^{\text{V}}(\omega) \right]^{2}$ in Eq.~(\ref{eq:ancilla}) which is a small quantity in the low-gain regime. Due to lack of phase-matching for the skewed parameter set, this term is even smaller leading to unnoticeable effects. These results indicate that in the low-gain regime, the ancilla intensity may not be the optimal quantity to measure to infer idler loss. Furthermore, the ancilla intensity is independent of additional phases and can not be used to construct interferograms.

    In the high-gain regime, Figs.~\ref{fig:ancilla}(b) and (d), we see that the normalized ancilla intensity and the corresponding normalized idler intensity are almost identical. Indeed, by looking at Eqs.~(\ref{eq:ic_idler}) and (\ref{eq:ancilla}) we see that both intensities approach $\eta_{I}(-\omega)\left[ N^{\text{V}}(\omega) \right]^{2}$ in the high-gain regime. Therefore, in the high-gain regime, the ancilla intensity mimics the idler intensity except that it operates at signal frequencies which are easier to detect experimentally. In the high-gain regime, one could in some sense ``directly'' infer the idler intensity and therefore incurred loss from the ancilla intensity. This is a clear strength of the IC interferometer.

    \subsubsection{Bleamsplitter arm intensity}\label{sec:arm}

\begin{figure}[ht]
	 	     \includegraphics[width=1\linewidth]{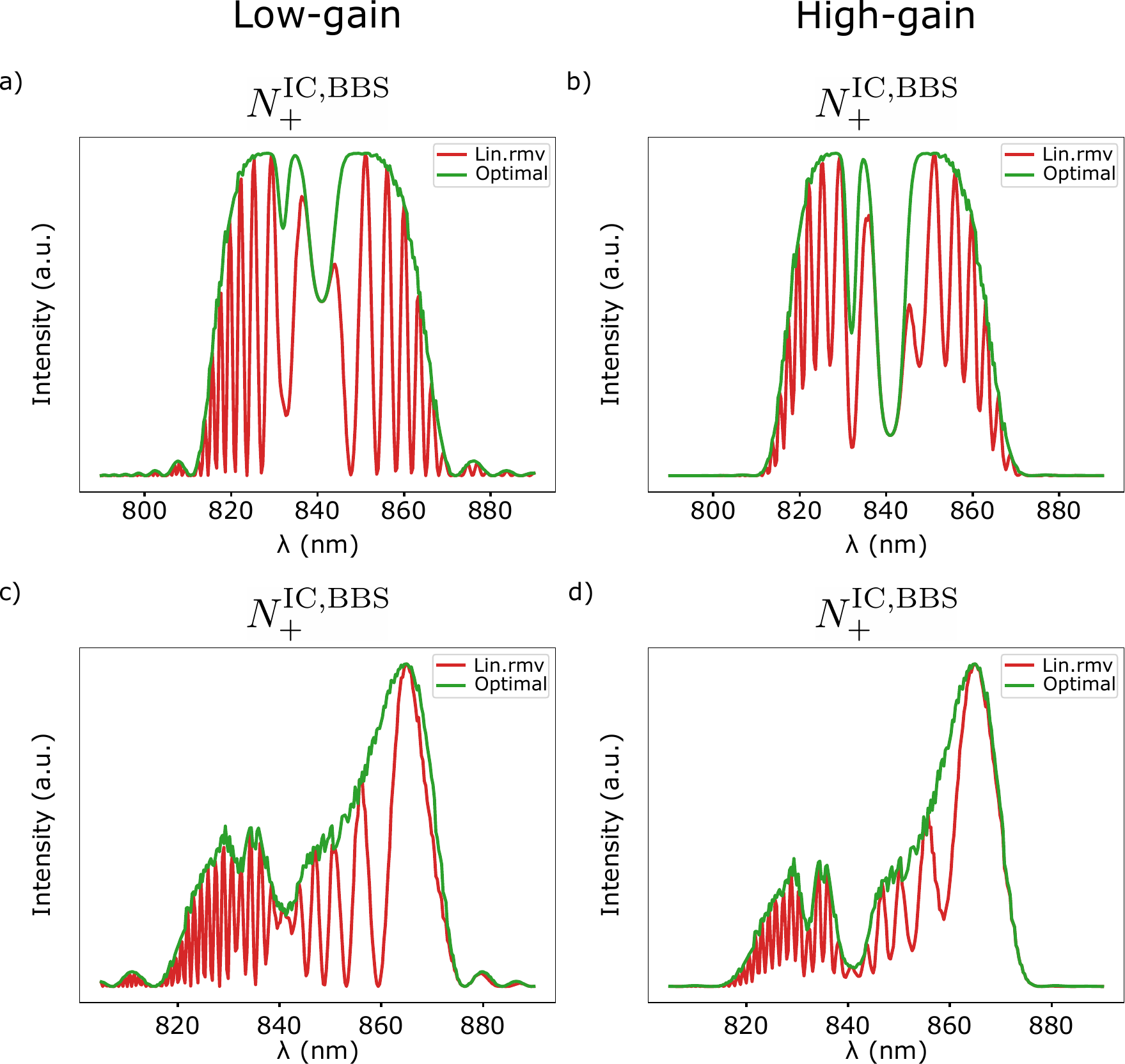}
	 	     \caption{Normalized intensity for the beamsplitter arm ($N^{\text{IC,BBS}}_{+}$) of the IC interferometer where additional idler phases were chosen to cancel the linear contribution in the $\Delta K(\omega)L/2$ term (c.f. Sec.~\ref{sec:IC}). Flat parameter set in the (a) low- and (b) high-gain regimes. Skewed parameter set in the (c) low- and (d) high-gain regimes. Green optimal curves represent the situation where the $\Delta K(\omega)L/2$ term is completely negated.}
	 	\label{fig:bs_intensity}
	   \end{figure}

    For the normalized beamsplitter arm intensity we consider $N^{\text{IC,BBS}}_{+}(\omega)$ (Eq.~(\ref{eq:IC_bbs})). In Fig.~\ref{fig:bs_intensity}, we show the normalized intensity for the flat parameter set in the (a) low- and (b) high-gain regimes and the skewed parameter set in the (c) low- and (d) high-gain regimes. As mentioned previously, we have chosen the additional phases $\Phi(\omega)$ such that the linear contribution of $\Delta K(\omega)L/2$ is cancelled out. Furthermore, for all four plots, we superpose an optimal curve where we completely negate the $\Delta K(\omega)L/2$ contribution. 
    
    Although the quadratic oscillations are present, we still see clear indications of the idler loss effects for both parameter sets in both the low- and high-gain limits. For the flat parameters, however, the effects are not as apparent due to said oscillations. Indeed, in the low-gain regime the effects of the weaker idler absorption are almost unnoticeable (the only indicator being the fact that the oscillation minima does not reach zero intensity compared to the others). This is not the case for the skewed parameter set where the oscillations are much faster at the absorption points, allowing one to more readily see the valleys. As expected from the theory, in the high-gain regime we also see that the intensity no longer vanishes and  we see that the high-loss idler signature becomes more noticeable. Due to the different observed behaviours, it is therefore important to consider the effects of these oscillations on a case by case basis for these intensity measurements (as shown in Ref.~\cite{Roeder_OCT} it is possible to construct sources with reasonable oscillations).
    
    Although these oscillations are somewhat detrimental for snap-shot intensity measurements, we can still obtain reliable information concerning idler loss when we consider interferograms that scan over an additional idler OPD (see Sec.~\ref{sec:interferometry}).

    \subsection{DL with idler absorption}\label{sec:dist_lost}
    We now consider the different effects of idler-only loss on the signal intensity for the DL model. Since this implementation of loss is more complex than a beamsplitter interaction implementation, we produce a more thorough analysis which we break down into low-gain and high-gain effects. Additionally, in Appendix~\ref{sec:losslevel} we compare the levels of overall loss experienced by the idler between the DL and beamsplitter loss models to reiterate an interesting result.

    \subsubsection{Low-gain effects}\label{sec:losseffect}

\begin{figure}[ht]
		\includegraphics[width=1\linewidth]{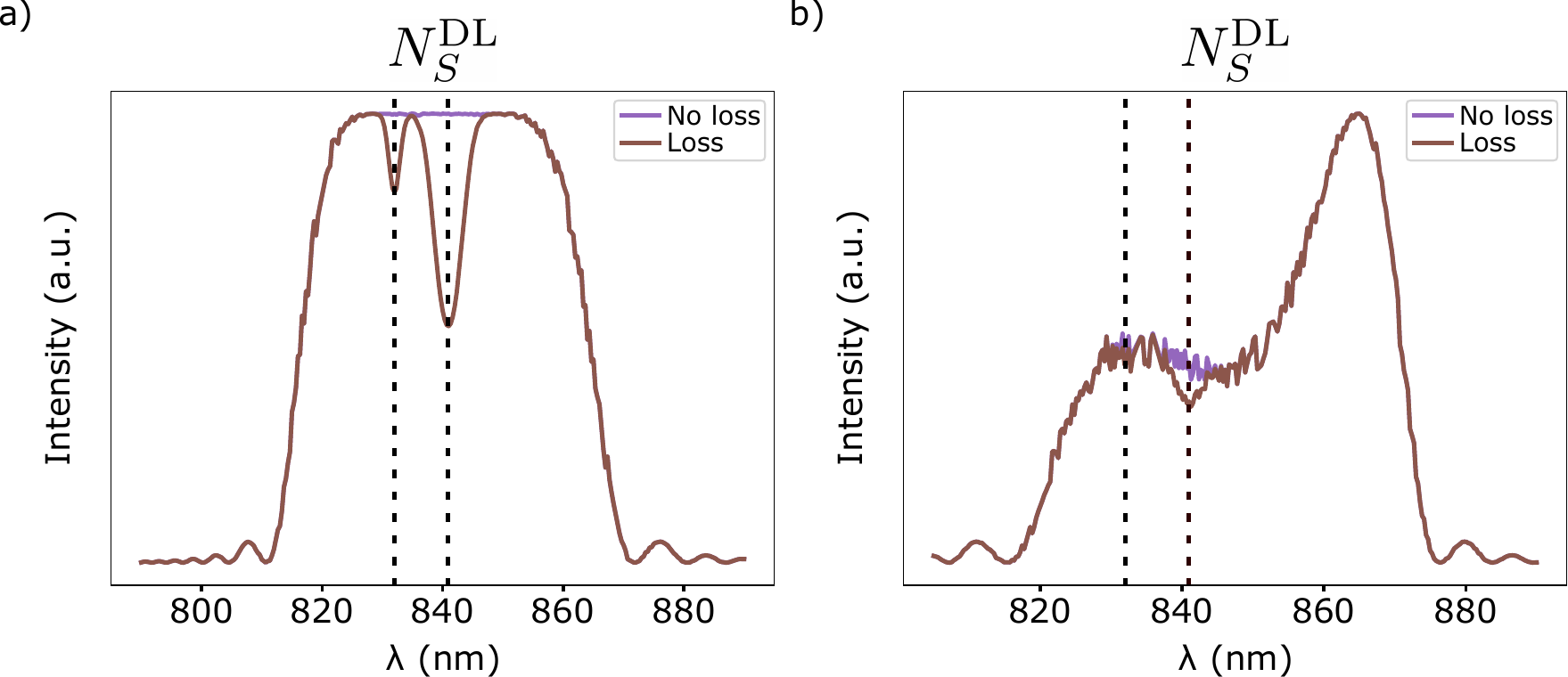}
		\caption{Effects of idler loss on signal intensity for the DL model. We compare the lossy and lossless signal intensities for the (a) flat parameter set and (b) skewed parameter set. Black dotted lines indicate the corresponding signal wavelengths of the idler absorption peaks. We see that even though we take the same decay rate for both sets of parameters, the effects of loss are much less noticeable for the skewed parameters (see Appendix~\ref{app:contrib} for discussion). Plots are shown for low-gain regime.}
		\label{fig:intensity}
	\end{figure}

    In Fig.~\ref{fig:intensity} we show how the normalized signal intensity is affected by idler loss for the (a) flat and (b) skewed parameter sets in the low-gain regime. Even though the signal has no intrinsic decay, it does experience loss due to the idler absorption.
    Furthermore, we see that the effects of idler loss differ greatly depending on the chosen set of parameters. For the skewed parameter set in Fig.~\ref{fig:intensity}(b) the effects on the signal intensity are much smaller and one does not observe a loss signature at a signal wavelength of $832\,\mathrm{nm}$, and, consequently, would conclude that there was no idler absorption at the corresponding wavelength of $2850\,\mathrm{nm}$. We emphasize again that we set losses of $70\%$ at $2850\,\mathrm{nm}$ and $99\%$ at $2750\,\mathrm{nm}$, see Tab. \ref{tab:absorption_parameters}. In Appendix~\ref{app:contrib}, we go over the different contributions to the signal intensity to show why both parameter sets give rise to different behaviours.
    
    This dependence on the initial signal intensity curve when considering idler loss effects puts into question how useful a single source could be for sensing and estimating idler absorption, especially when considering the experimental challenges in meeting target operation conditions for highly engineered sources.

	\subsubsection{High-gain effects}\label{app:highgain}

\begin{figure}[ht]
		\includegraphics[width=1\linewidth]{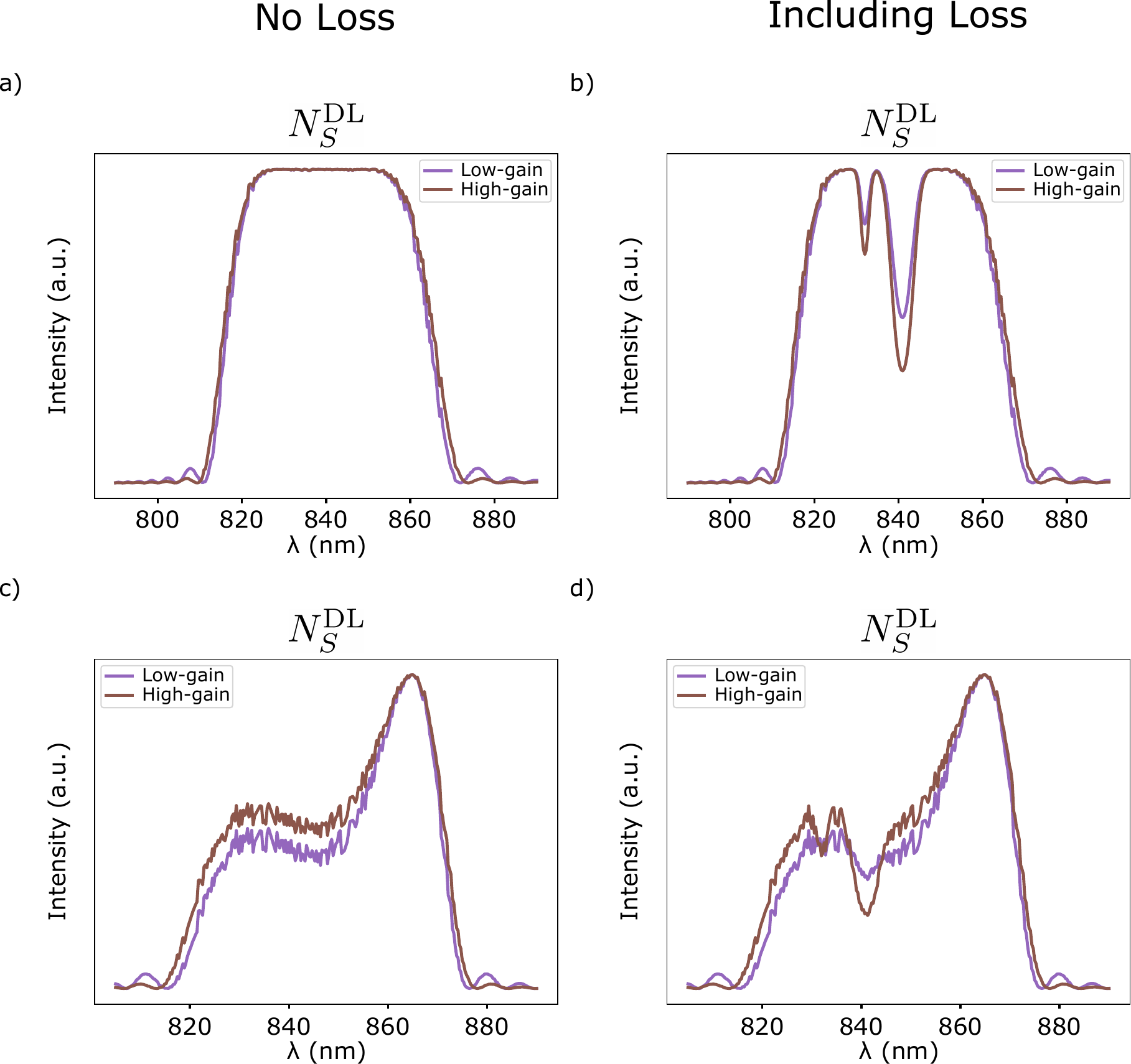}
		\caption{Normalized signal intensity for low- and high-gain regimes. We compare low- and high-gain intensities for the flat parameters (a) without loss and (b) with loss. We compare low- and high-gain intensities for the skewed parameters (c) without loss and (d) with loss. For low(high)-gain regimes we have $N^{P}_{S}=0.04(14)$. For the lossy cases, higher levels of gain lead to more noticeable effects from the idler loss.}
		\label{fig:high_gain}
	\end{figure}

    In Fig.~\ref{fig:high_gain}(a), we show the signal intensity for the flat parameter set without loss in both low- and high-gain regimes. The normalized curves are very similar for both levels of gain. The high-gain curve has a slightly larger bandwidth, due to a small shift in the zeros of the intensity (see Fig.~\ref{fig:vacuum}(a) to see how the zeros shift). For the source configuration considered here, the broadening is small. Increasing the interaction strength increases the value of the PMA for which the signal intensity goes to zero and therefore increases the range of frequencies for which amplification occurs. However, due to the cubic nature of the PMA this increase leads to a small variation in wavelength. 

    When we include loss, we see that as we increase gain, the effects of idler loss on the signal intensity become more pronounced as shown in Fig.~\ref{fig:high_gain}(b). The reason for this effect is two-fold: both the bare and added contributions are responsible for this enhancement. We can understand this effect by considering the dominating exponential term from both contributions. As we are well inside the plateau region, we have that $\Sigma_{K}\approx 0$ and the normalization amounts to a division by $\sinh^{2}(\gamma L)$. In this parameter regime, we can do the added noise integral and we find that both contributions have the same dominating exponential which takes the form $\exp\left[\left( \sqrt{4\gamma^{2}+(\kappa_{I}/2)^2}-2\gamma-\kappa_{I}/2\right)L\right]\equiv \exp\left[ A(\gamma)L \right]$. To show that the exponential decreases as a function of gain, we consider the derivative of the argument w.r.t $\gamma$
     \begin{align}
         \frac{\partial A(\gamma) }{\partial \gamma} &= 2\left[ \frac{2\gamma}{\sqrt{4\gamma^{2}+(\kappa_{I}/2)^2}} -1 \right]\nonumber\\
         & = 2\left[ \frac{1}{\sqrt{1+(\kappa_{I}/4\gamma)^2}} -1 \right].
     \end{align}
     For a fixed level of loss, we have that the derivative is negative since the inverse square-root term is less than one. In fact, the derivative is always negative except in the large-$\gamma$ limit where it tends to zero from the left at which point the normalized signal intensity tends toward $\exp[-\kappa_{I}L/2]$. Since the derivative is negative, the argument and therefore the exponential decrease as a function of gain. For the added noise contribution, we also have a $\gamma$-dependent proportionality factor which takes the form $\tfrac{\kappa_{I}}{4}(\kappa_{I}+2\sqrt{4\gamma^{2}+(\kappa_{I}/2)^2})/(4\gamma^{2}+(\kappa_{I}/2))$. This factor also decreases as a function of $\gamma$ and so the added noise contribution decreases even more as a function of $\gamma$. Therefore, for the same level of loss, we expect both contributions and therefore the intensity to decrease as we increase gain.

    In Fig.~\ref{fig:high_gain}(c) we show the signal intensity for the skewed parameter regime, without loss, in the low- and high-gain regimes. As we can see, in the high-gain regime, the relative height of the plateau region compared to the peak increases. This overall increase is due to the fact that as we increase the nonlinear gain $\gamma$ we went from a regime where $N_{S}\propto \sin(\nu(\omega)L/2)$ to $N_{S}\propto \sinh(\nu(\omega)L/2)$ and so the plateau also started to experience exponential growth. This behaviour is again dependent on the dispersion relations used. Similarly to the flat parameter set, we also see a small broadening of the amplification region.

    Finally, in Fig.~\ref{fig:high_gain}(d) we show the normalized signal intensity for the skewed parameter set, including loss, in the low- and high-gain regimes. We now have a combination of the two effects described above: the relative height of the plateau increases and the effects of idler only loss are more pronounced. In contrast to Fig.~\ref{fig:intensity}, we now observe both idler absorption features. We conclude that, while high-gain effects increase the impact of idler loss on the signal intensity, it might not be possible to reach those levels experimentally and the performance of the distributed loss model will depend on the operation point of the source.

    \section{Additional dispersion effects}\label{sec:dispersion}

    Having explored the effects of idler-only loss without additional dispersion, we now consider the opposite combination where we allow for additional idler dispersion without any idler loss.  Before combining both effects, it is important to understand each on its own. Furthermore, this situation is also relevant in the interferometric configurations when wanting to determine the additional dispersion accrued by the idler via an unknown analyte\cite{roeder2024su11prx, Riazi2019dispersion}. For the DL model we consider a different effect where-upon the analyte induces small variations in the idler dispersion without inducing any loss.

	\subsection{SU(1,1) with idler OPD}
	\label{sec:su11_opd}
    
    As previously proposed and demonstrated (see, e.g., \cite{roeder2024su11prx, Riazi2019dispersion}), one can use frequency oscillations in the signal intensity of an SU(1,1) interferometer to determine the level of additional dispersion accrued by the idler in between the nonlinear regions. 
    
    In fact, it can be shown that in the low-gain regime ($N^P_S\ll1$) the normalized oscillations in the signal intensity as a function of frequency behave as~\cite{Riazi2019dispersion}
    \begin{align}\label{eq:osc_low}
        N^{\text{osc.low}}_{S}(\omega) \propto \cos{\left[ \Phi(\omega)+\Sigma_{K}(\omega)L  \right]}
    \end{align}
    where  $\Sigma_{K}(\omega)L$ originates from the waveguide dispersion of the squeezers. By initially characterizing the squeezers, we can then infer the additional dispersion in the low-gain regime. 

    From the SU(1,1) solutions above, we can now obtain an expression for the frequency oscillations in the signal intensity which are valid for all levels of gain. The only oscillatory term in Eq.~(\ref{eq:ns_su11}) stems from the last term. Isolating this term and normalizing, we find that the oscillations are
    \begin{align}\label{eq:osc_full}
        N^{\text{osc.gen}}_{S}(\omega) \propto \cos{\left( \Phi(\omega) +\Psi(\omega)  \right)}
    \end{align}
    where $\Psi(\omega)$ is given in Eq.~(\ref{eq:psi_full}). Note that in the following analysis we ignore the constant $+\pi$ as it only modifies the sign of the cosine.
    
    Since the gain dependence of Eq.~(\ref{eq:psi_full}) is not intuitive, we expand $\Psi(\omega)$ as a function of the PMA $\Sigma_{K}(\omega)$ (since in the amplification plateau we are close to perfect phase-matching and  $\Sigma_{K}\approx 0$) to obtain an approximation with explicit gain dependence:
    \begin{align}\label{eq:psi_approx}
        \Psi(\omega)\approx& \frac{\tanh{\gamma L}}{\gamma}\Sigma_{K}(\omega)+\mathcal{O}(\Sigma_{K}(\omega)^{3}).
    \end{align}
    In the low-gain regime ($\gamma\ll1$), $\tanh{(\gamma L)} / \gamma \approx L$ and we obtain Eq.~(\ref{eq:osc_low}).
    For the following analysis, we refer to the approximation where we use Eq.~(\ref{eq:psi_approx}) in Eq.~(\ref{eq:osc_full}) as the ``gain-dependent approximation''.
    
\begin{figure}[t]
		\includegraphics[width=1\linewidth]{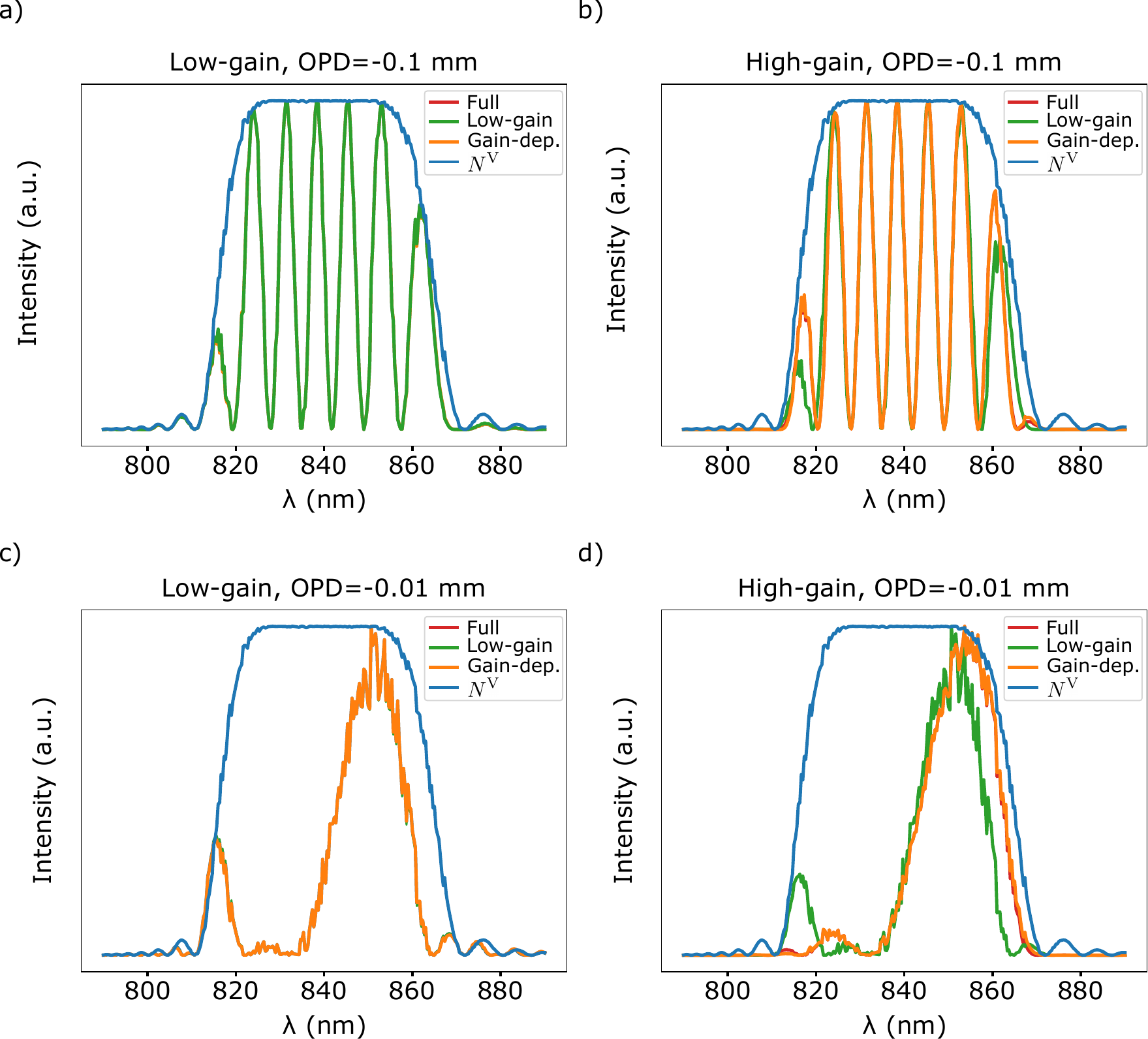}
		\caption{Normalized signal intensity for the SU(1,1) model with idler optical path delays (OPDs) and without any additional dispersion or loss for the flat parameters. (a) and (b) represent the normalized signal intensities for the low- and high-gain regimes, respectively, with an OPD of $-0.1$ mm. (c) and (d) represent the normalized signal intensities for the low- and high- gain regimes, respectively, with an OPD of $-0.01$ mm. The red curve is the full analytic solution (cf. Eq.~\ref{eq:ns_su11} of main text), the orange curve is the low-gain solution where the oscillations take the form of Eq.~\ref{eq:osc_low}, the green curve is the gain-dependent approximation described in Sec.~\ref{sec:su11_opd}, and the blue curve is the normalized signal intensity for a lossless single pass that we include to show the amplification region. The gain-dependent approximation follows the full solution quite well.}
		\label{fig:opd_gain}
	\end{figure}

    In Figs.~\ref{fig:opd_gain}(a) and (b) we show the signal intensity pattern for an SU(1,1) interferometer with an additional idler OPD of $-0.1$ mm and without any additional dispersion or loss in the low- and high-gain regimes, respectively.
    The oscillations are very well defined inside the amplification plateau. As a function of gain, they vary minimally and Eq.~(\ref{eq:osc_low}) holds to a good approximation. It only deviates as we near the edges of the plateau in the high-gain regime, where, due to the increasing phase mismatch, the overall intensity decays quickly to zero. Here, we find that the gain-dependent approximation gives better agreement with the full solution.
    
    Choosing different OPDs and additional dispersion will change how well the low-gain approximation of Eq.~(\ref{eq:osc_low}) holds. To show this, in Figs.~\ref{fig:opd_gain}(c) and (d) we show the signal intensity for the same levels of gain but a smaller OPD of $-0.01$ mm. We see that the low-gain approximation differs greatly in the high-gain regime, to the point where oscillation maxima are predicted at wrong wavelengths. Again, the gain-dependent approximation yields predictions that are much closer to the full solution. In Appendix~\ref{app:skewed_opd}, we include a similar figure and brief discussion of the same analysis for the skewed parameter set.

    In the first case, the OPD is much larger and is the dominating term inside the cosine. As such, the oscillations are less sensitive to high-gain effects. In the second case, the OPD is smaller making the intensity more sensitive to high-gain effects. While for some measurements one aims to maximize the number of spectral fringes and is thus insensitive to gain dependent effects \cite{Riazi2019dispersion}, other measurements such as time-domain optical coherence tomography or Fourier-transform infrared spectroscopy may need to take into account these effects to achieve maximum measurement performance.
    
	\subsection{IC with idler OPD}
	\label{sec:ic_opd}
    
    For the IC configuration, we find that the normalized beamsplitter arm oscillations are 
    \begin{align}
       N^{\text{osc.gen}}_{S}(\omega) \propto \sin(\frac{\Delta K(\omega) L}{2}-\Phi(\omega)-\Psi(\omega)/2)
    \end{align}
    and, as mentioned previously, contain an additional phase term $\Delta K(\omega)L/2$. Regardless of whether or not we are phase-matched, this additional phase term is present and needs to be considered if one wants to infer the additional dispersion accrued by the idler mode. The $\Psi(\omega)$ contribution can be treated the same way as the SU(1,1) configuration. Furthermore, we consider the experimental setup where we configure the free-space arm propagation lengths to cancel the linear contribution of the additional phase term. We then add an additional idler only OPD and determine the effects of said OPD over the remaining contributions of $\Delta K(\omega)L/2$. 
    
    For the source considered in this manuscript, we find that the quadratic contribution of the additional phase dominates over the chosen idler OPDs of -0.1 and -0.01 mm and these OPDs do not lead to any noticeable effects when considering the beamsplitter arm intensity (i.e. the oscillations in the intensity curves are the same as Figs.~\ref{fig:ic_oscillation}(b) and (d) modulo a constant phase-shift). As such, we do not include figures including the idler only OPDs for the IC configuration. Note that the additional phase term is gain-independent and therefore any high-gain effects will come from the $\Psi(\omega)$ term which can be treated as is done for the SU(1,1) configuration. Again, we stress that this result is specific to the dispersion engineered source used in this manuscript.

	\subsection{DL with anomalous dispersion}
	\label{sec:dist_anom}

    For the DL model, we imagine the case where the analyte induces changes in the dispersive properties of the idler. We can think of the analyte as being some foreign cladding over the waveguide but only at certain frequencies. Therefore, at these frequencies, the presence of the analyte modifies the idler mode profile and effective dispersion in the waveguide. At the level of the Sellmeier equations used to generate the dispersive data, these effects would also not be included and so here we include them manually. Note that when deriving the DL model equations of motion, the signal and idler dispersions can be arbitrarily chosen and so simply adding an anomalous dispersion term for the idler is permissible. 
    
    To this end, we allow for small variations in the idler dispersion relations such that  
    \begin{align}
        \Delta k_{I}(\omega)\rightarrow \Delta k_{I}(\omega)-0.1\kappa_{I}(\omega)
    \end{align}
    where $\kappa_{I}$ is indeed the decay rate (described in Sec.~\ref{sec:loss}) but here it comes without the factor of $i$ and as such modifies the dispersion. We choose weak variations because any sizable effect will lead to complete phase-mismatch and very apparent bands without amplification. We also choose $\kappa_{I}(\omega)$ such that the anomalous dispersion turns on/off in the same fashion that the absorption does.

\begin{figure}[ht]
		\includegraphics[width=1\linewidth]{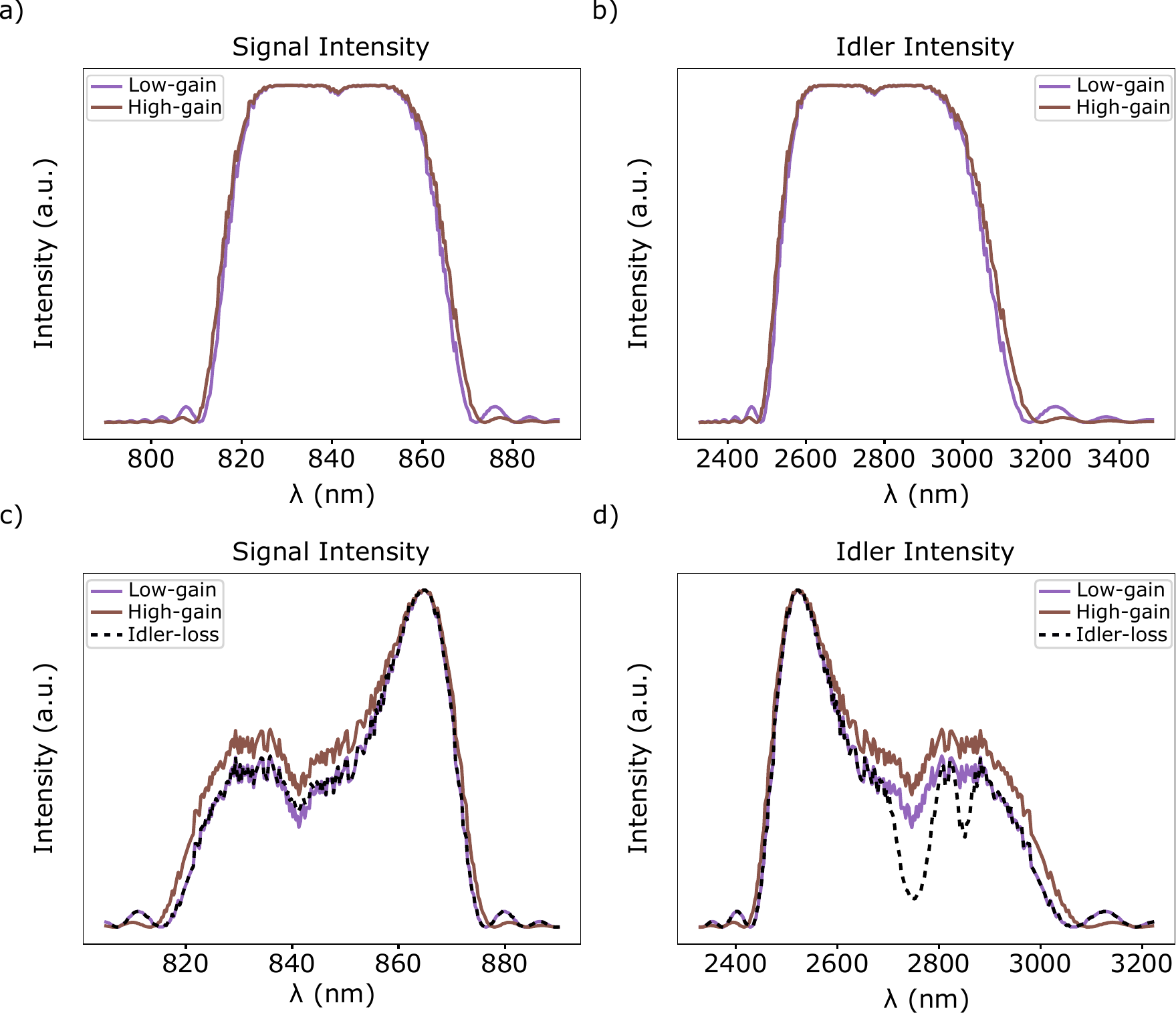}
		\caption{Normalized intensities for the DL model with anomalous dispersion (c.f. Sec.~\ref{sec:dist_anom}). (a) signal intensity and (b) idler intensity for the flat parameters in the low- and high-gain regimes. (c) signal intensity and (d) idler intensity for the skewed parameters in the low- and high-gain regimes. The anomalous dispersion signature is the same for signal and idler intensities. Furthermore, this signature is gain independent. For the skewed parameter set, we overlay the low-gain idler loss intensities without anomalous dispersion(black-dotted curve).}
		\label{fig:anomdisp}
	\end{figure}

    In Fig.~\ref{fig:anomdisp} we show the normalized (a) signal and (b) idler intensities in the low- and high-gain for the flat parameter set. In this case, we see that the anomalous dispersion effects are quite small. In panels (c) and (d) we show the signal and idler intensities, respectively, for the skewed parameter set. In this case, we see that the anomalous dispersion term gives rise to a much greater effect which looks similar to the low-gain idler loss effect shown in Fig.~\ref{fig:high_gain}(d) (black-dotted curve in Fig.~\ref{fig:anomdisp}(c)). However, unlike the idler loss signature, the anomalous dispersion effects are gain independent: higher levels of gain do not lead to shallower valleys in intensity. Furthermore, the effects of the anomalous dispersion are the same for the signal and idler intensities: the idler intensities are perfect mirrors of the signal intensities. The discrepancy between the effects of anomalous dispersion on the two parameter sets can be attributed to the fact that for the skewed parameter set we are already phase-mismatched and the addition of this term makes things more mismatched. Indeed, for the skewed parameter set, this anomalous dispersion could change the signal intensity behaviour from sinh-like to sinc-like and therefore suppress exponential growth. 
    
    For optimal sources, idler loss and anomalous dispersion effects can be readily differentiated, even without access to the idler intensity. However, for situations where we can not measure the idler and the source is non-optimal the effects can be quite similar. It is therefore important to verify the gain-dependence of the signal intensity valleys to determine if they are caused by anomalous dispersion or absorption (and possibly both).

    \section{Interferometry}\label{sec:interferometry}

    With both idler-only effects studied individually, we now study configurations where both effects are present. In interferometric systems, it is common to scan the desired intensity over a range of OPDs. This allows one to span the whole region of operation for the second pass: going from perfect squeezer to perfect anti-squeezer for each frequency in the range of relevant frequencies. From these scans, we can study the full 2-dimensional interferogram as well as the visibility which allows one to gain information concerning the idler loss. Unlike the SU(1,1) and IC configurations, the DL model does not on its own allow for such interferometric measurements. However, for the sake of comparison, we consider a different SU(1,1) configuration where we use DL sources, rather than lossless ones, without additional loss but allowing for idler-only OPDs in between the lossy sources. The nonlinear update rules can be readily applied to DL sources, however, the analytics get rather convoluted. As such, we only produce numerical scans for the DL-SU(1,1).

	\subsection{Spectral interferogram}
	\label{sec:interferogram}

\begin{figure*}[ht]
	 	\includegraphics[width=1\linewidth]{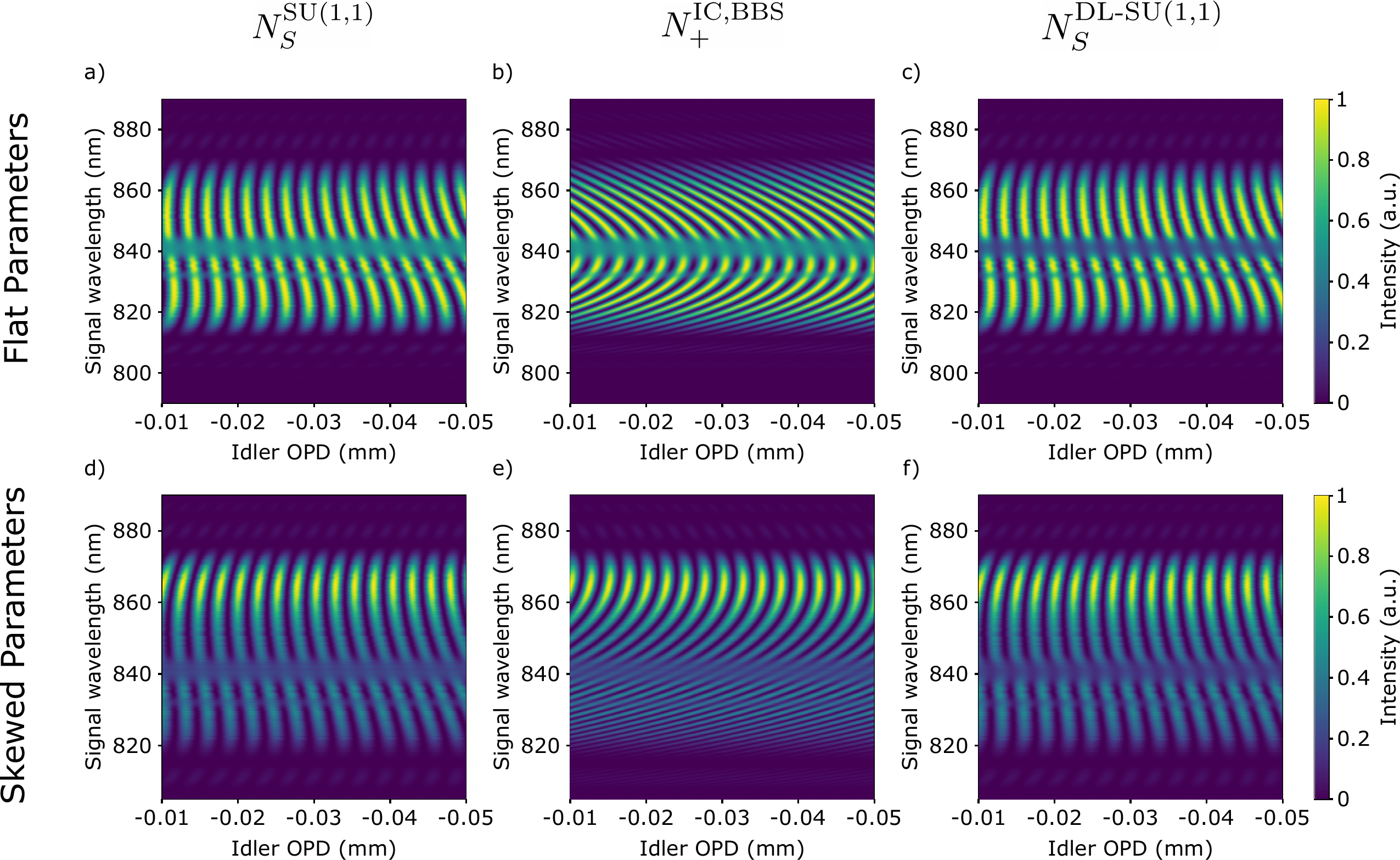}
	 	\caption{Spectral interferograms for SU(1,1), DL-SU(1,1), and IC operation. We plot the normalized intensities as a function of signal wavelength and idler optical path delay (OPD). The idler loss is implemented as described in Sec.~\ref{sec:loss}. Spectral interferogram for flat parameter set for (a) SU(1,1), (b) IC, and (c) distributed DL-SU(1,1) operation. Spectral interferogram for skewed parameter set for (d) SU(1,1), (e) IC, and (f) distributed DL-SU(1,1) operation. For the DL-SU(1,1) and SU(1,1) configurations, we plot the signal intensity. For the IC configuration, we plot the beamsplitter arm intensity. For the DL-SU(1,1) operation, we use DL sources without loss in between said sources.} 
	 	\label{fig:interferometer}
	 \end{figure*}

    We construct 2D spectral interferograms by sweeping over a range of idler OPDs and recording the desired intensity at each point.  We vary the idler OPD from $-0.01$ to $-0.05$ mm while keeping the level of gain fixed. For plotting purposes, we only consider the low-gain regime where loss effects were originally unclear (i.e. for the IC and DL configurations). In Fig.~\ref{fig:interferometer} we show several 2D spectrograms:  (a) and (d) are the SU(1,1) signal spectrograms for the flat and skewed parameter sets, respectively; (b) and (e) are the IC beamsplitter arm spectrograms for the flat and skewed parameter sets, respectively; (c) and (f) are the  DL-SU(1,1) signal spectrograms for the flat and skewed parameter sets, respectively. 
    For the SU(1,1) and DL-SU(1,1) interferograms we see clear and distinct lines of approximately constant intensity where the idler absorption occurs for both sets of parameters. The contrast of the bands is not as strong for the skewed parameter set since the original level of gain is lower in the skewed plateau. We also see that the broader absorption peak is more apparent. For the DL-SU(1,1), even though the idler loss signatures on the signal intensity are barely noticeable in the low-gain, there are clear bands in the interferogram. Hence, we can still gather idler loss information from the distributed loss source signal intensity, albeit working in an SU(1,1) setup, for the skewed parameters in the low-gain regime. Note that the phase of the fringes in the DL-SU(1,1) in Fig.~\ref{fig:interferometer}(f) is affected by the absorption feature; this is to be expected due to the effects of loss matching~\cite{antonosyan2014effect,helt2015spontaneous,helt2020degenerate} in the PMF in Eq.~\eqref{eq:lossmatching}.

    For the IC configuration, we again consider the case where we cancel the linear oscillations in the beamsplitter arm intensity. The behaviour is very similar to the SU(1,1) and DL-SU(1,1) configurations, however, the skewed parameters bands are not as distinct. Due to the fact that we have quadratic oscillations present, the intensity bands are curved rather than being parallel to each other. This curvature is more accentuated in the skewed plateau region making it harder to see the absorption bands.

     \subsection{Visibility}\label{sec:vis}
     From the 2D interferograms, we define the spectral Visibility, $V(\omega)$, as
     \begin{align}\label{eq:visibility}
         V(\omega) = \frac{ \text{max}_{\text{OPD}}\left[N_{S}(\omega)\right]-\text{min}_{\text{OPD}}\left[N_{S}(\omega)\right]}{\text{max}_{\text{OPD}}\left[N_{S}(\omega)\right]+\text{min}_{\text{OPD}}\left[N_{S}(\omega)\right]}
     \end{align}
     where for each frequency we find the max and min values of the signal intensity w.r.t. the OPD. As we scan through the idler OPD, we have that the max(min) values occur when the oscillatory term in the intensity takes on the value 1(-1). Indeed, it is possible to re-write Eqs.~(\ref{eq:ns_su11}) and (\ref{eq:IC_bbs}) (as well as the signal DL-SU(1,1) intensity which we have omitted) in the form
     \begin{align}
         N^{\text{IF}}(\omega) = \alpha(\omega) + V(\omega)\cos(\phi)
     \end{align}
     where the superscript ``IF'' is to indicate ``interferometer'', $\alpha(\omega)$ is non-oscillatory, $\phi$ represents the relevant phases, and $V(\omega)$ is the visibility as defined above. Note that the visibility varies between 0(low) and 1(high).
     
\begin{figure}[ht]
	 	\includegraphics[width=1\linewidth]{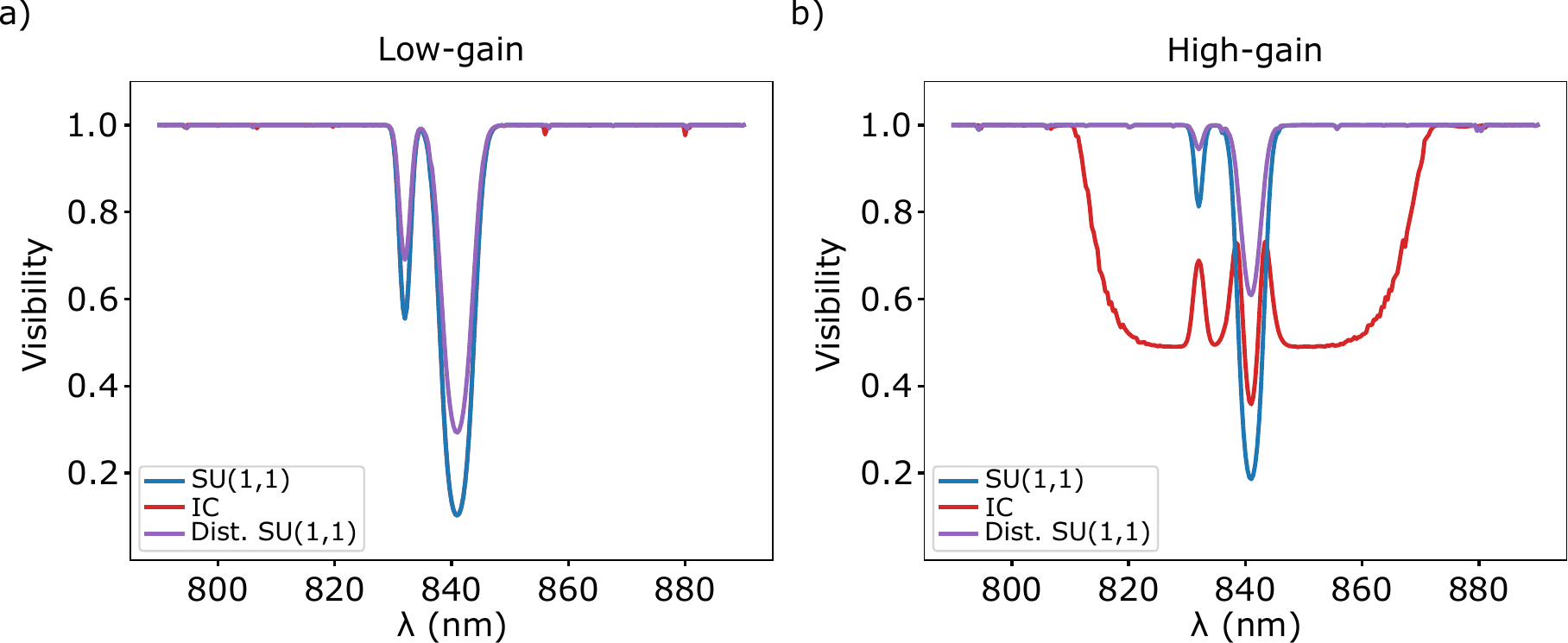}
	 	\caption{Comparing the visibilities of all three models in the (a) low-gain and (b) high-gain regimes. In the low-gain, the SU(1,1) and IC visibilities are very similar and the DL visibility is higher. In the high-gain the IC visibility differs from the SU(1,1) (see Sec.~\ref{sec:vis_analytics} for discussion) and all three visibilities increase. Plots shown for the flat parameters (skewed parameters gave similar results).}
	 	\label{fig:visbility}
	  \end{figure}

     In Fig.~\ref{fig:visbility}, we show the visibility for the three interferometric configurations in the (a) low- and (b) high-gain regimes for the flat parameter set. We do not include the skewed parameter set as the results were equivalent. Although we are not phase-matched for the skewed parameter set, by scanning over the idler OPD we also find the points of optimal squeezing/anti-squeezing resulting in the same visibility. The effects of idler loss are apparent for all configurations in both gain regimes.

     In the low-gain regime, the SU(1,1) and IC configurations have the same visibility. This can be explained analytically and will be addressed shortly. Comparatively, the visibility for the DL-SU(1,1) configuration is higher. From the definition, we find that the numerator in Eq.~(\ref{eq:visibility}) behaves very similarly for all three configurations (i.e. the difference between max and min values is very similar) and so the discrepancy in the visibility comes from the denominator. For the DL-SU(1,1) model, since both sources are lossy, the loss is applied twice which in turn causes the max and min values, at the corresponding absorption peaks, to be lower than in the SU(1,1) model. This in turn leads higher visibilities for the DL-SU(1,1) model.   

     The high-gain regime visibilities of the SU(1,1) and DL-SU(1,1) configurations behave similarly to the low-gain regime but with higher visibilities. The discrepancy between the low- and high-gain DL-SU(1,1) visibilities is bigger than that of the SU(1,1) visibilities. This is again attributed to the compounded loss in the DL-SU(1,1) configuration. These behaviours are in sharp contrast with the IC visibility which behaves quite distinctly in the high-gain regime while still retaining clear signatures of idler-only loss.

    \subsubsection{SU(1,1) and IC analytics}\label{sec:vis_analytics}
    For the SU(1,1) and IC models, we can obtain analytic expressions for the visibility which allows us to look at the low- and high-gain behaviour more readily. When idler loss is present, we find that
    \begin{align}
        V^{\text{SU(1,1)}}(\omega) &= \frac{2\sqrt{\eta_{I}(-\omega)}\left( N^{\text{V}}(\omega) +1 \right)}{2+N^{\text{V}}(\omega)\left(   1+\eta_{I}(-\omega)\right)},\\
         V^{\text{IC}}(\omega) &= \frac{2\sqrt{\eta_{I}(-\omega)\left( N^{\text{V}}(\omega) +1 \right)}}{2+\eta_{I}(-\omega)N^{\text{V}}(\omega)}.
    \end{align}
    Without any loss, $\eta_{I}(-\omega)=1$, we see that $V^{\text{SU(1,1)}}(\omega)=1$ while $V^{\text{IC}}(\omega)$ remains a function of $N^{\text{V}}$. This is because in the IC model, we can never truly anti-squeeze. For the SU(1,1) visibility, we find that the derivative w.r.t. $N^{\text{V}}(\omega)$ is greater than zero for non-zero loss, which tells us that the visibility increases as a function of gain.
    
    In the low-gain limit, we have that $N^{\text{V}}(\omega)\ll 1$ giving us
    \begin{align}
        V^{\text{SU(1,1)}}(\omega)\approx V^{\text{IC}}(\omega)\approx \sqrt{\eta_{I}(-\omega)}
    \end{align}
    which agrees which the results of Fig.~\ref{fig:visbility}(a) (the visibility minima correspond to $0.56\approx\sqrt{0.3}$ and $0.101\approx\sqrt{0.01}$). Note that in the low-gain limit, we do find that the IC visibility is approximately unity without loss.

    For the high-gain limit, one needs to be careful when taking limits. Although the mathematical limit $N^{\text{V}}(\omega)\gg 1$ can give nice limiting behaviour, it should be interpreted with caution. Conventionally, the high-gain limit in quantum optics refers to $N^{\text{V}}(\omega)\gtrapprox 1$. Recall that we chose $N^{P}_{S}=14$ as the high-gain threshold. Furthermore, the high-gain limit is only valid in the plateau region and does not properly explain the transition between the phase-matched and non-matched regions. Regardless, by Taylor expanding the visibilities as a function of $1/N^{\text{V}}(\omega)$ we obtain  expressions valid in the plateau regions for the SU(1,1) and IC configurations in the high-gain limit
    \begin{align}
        V^{\text{SU(1,1)}}(\omega)\approx\frac{\sqrt{\eta_{I}(-\omega)}}{1+\eta_{I}(-\omega)},\\
        V^{\text{IC}}(\omega)\approx \frac{2}{\sqrt{\eta_{I}(-\omega)N^{\text{V}}(\omega)}}.
    \end{align}
    For the high-gain regime of the SU(1,1) configuration we find the visibility minima to be 0.79 and 0.177 rather than the limit values of 0.843 and 0.198. For the IC configuration, the inverse dependence on $\sqrt{N^{\text{V}}(\omega)}$ helps explain the behaviour of Fig.~\ref{fig:visbility}(b). However, for the chosen level of $N^{P}_{S}$, we are not high-gain enough to see the full limiting behaviour, especially at the frequencies where idler loss is present. In the mathematical high-gain limit, we would expect the visibility to drastically drop to zero and show peaks at the corresponding idler loss frequencies for the IC configuration.

    From this analysis, we conclude that in terms of spectrograms and visibility, the low-gain regime is optimal for inferring idler-only loss.

	\section{Conclusion}
	\label{sec:conc}

    In this manuscript, we have studied the outcomes of low- and high-gain unmeasured idler-only effects on the measured signal spectra for highly engineered broadband PDC sources in the CW limit and for different sensing configurations. We considered the SU(1,1), IC, and DL sensing configurations as well as two operation points for the sources: the flat (ideal) and skewed (non-optimal) parameter sets. 

    By first considering the effects of idler-only loss, we show that higher levels of gain leads to more prominent signatures on the signal intensity for all sensing configurations. For the SU(1,1) configuration, we find that because of phase-mismatch the skewed parameter set leads to gain dependent operation of the second squeezer. In the high-gain limit, we show that for the IC configuration the ancilla intensity approaches the behaviour of the idler intensity, effectively allowing one to probe the idler intensity at signal wavelengths. However, when recombining the ancilla and the signal, we find additional oscillations which are detrimental to observations of the loss signatures. Although we can cancel the oscillations due to group velocity matching, the group-velocity-dispersion terms are still present and in fact bolstered for highly engineered sources. For the DL configuration, we find that depending on the operation point of the source, the loss signatures can be unnoticeable in the low-gain due to added noise contributions. 

    We then studied the effects of idler-only additional dispersion without the presence of absorption. For the SU(1,1) and IC configurations, high-gain effects can lead to shifting oscillatory peak positions when compared to low-gain result. These shifts depend on the level of additional dispersion and so it is important to account for these high-gain effects when trying to determine the amount of additional dispersion from the oscillatory peak positions. For the DL configuration, we considered the case where the idler experiences additional small frequency dependent anomalous dispersion. We find that the signatures on the signal intensity are gain-independent and more apparent for the skewed parameter set due to phase-mismatch being already present. Indeed, for the skewed parameter set, the signatures of the anomalous dispersion are very similar to those of absorption. However, the gain-independence of the effect allows one to differentiate between the effects of anomalous dispersion and absorption.  

    Combining the effects of idler-only absorption and additional dispersion via variable OPDs, we studied 2D interferograms and the visibilities of the three sensing configurations. From the interferograms, we observe clear bands at the corresponding signal frequencies caused by idler-only absorption for most configurations. For the IC configuration, the additional quadratic term present in the oscillations leads to a blurring of bands for the skewed parameter set. In the low-gain limit, we find that the SU(1,1) and IC visibilities are the same. The DL visibility behaves similarly as well, but is higher due to the compounded effects of loss when considering an SU(1,1) interferometer with lossy sources. In the high-gain, we find that the SU(1,1) and DL visibilities increase whereas the IC visibility behaves unusually due to its dependence on the signal intensity. 

    These results are relevant for the proper analysis of experimental interferometric signal spectra. We point out specific effects which should be taken into account as well as how to differentiate them accordingly. Given the dispersion data of any source, one can follow our analysis to characterize their source and determine which configuration is best suited for their needs and operation points.

    \section*{Acknowledgements}
	MH and NQ acknowledge support from the Minist\`{e}re de l'\'{E}conomie et de l’Innovation du Qu\'{e}bec and the Natural Sciences and Engineering Research Council of Canada. This work has been funded by the European Union's Horizon Europe Research and Innovation Programme under agreement 101070700 project MIRAQLS. FR is member of the Max Planck School of Photonics supported by the German Federal Ministry of Research, Technology and Space (BMFTR), the Max Planck Society, and the Fraunhofer Society. 

    \section*{Data Availability Statement}
    
     Raw data and analysis code are available from the corresponding author upon reasonable request.

	\appendix

	\section{Signal intensity contributions for the distributed model}\label{app:contrib}
    In this appendix, we discuss why the two different parameter regimes lead to different effects on the signal intensity when only the idler mode experiences loss in the distributed loss model.

\begin{figure}[ht]
		\includegraphics[width=1\linewidth]{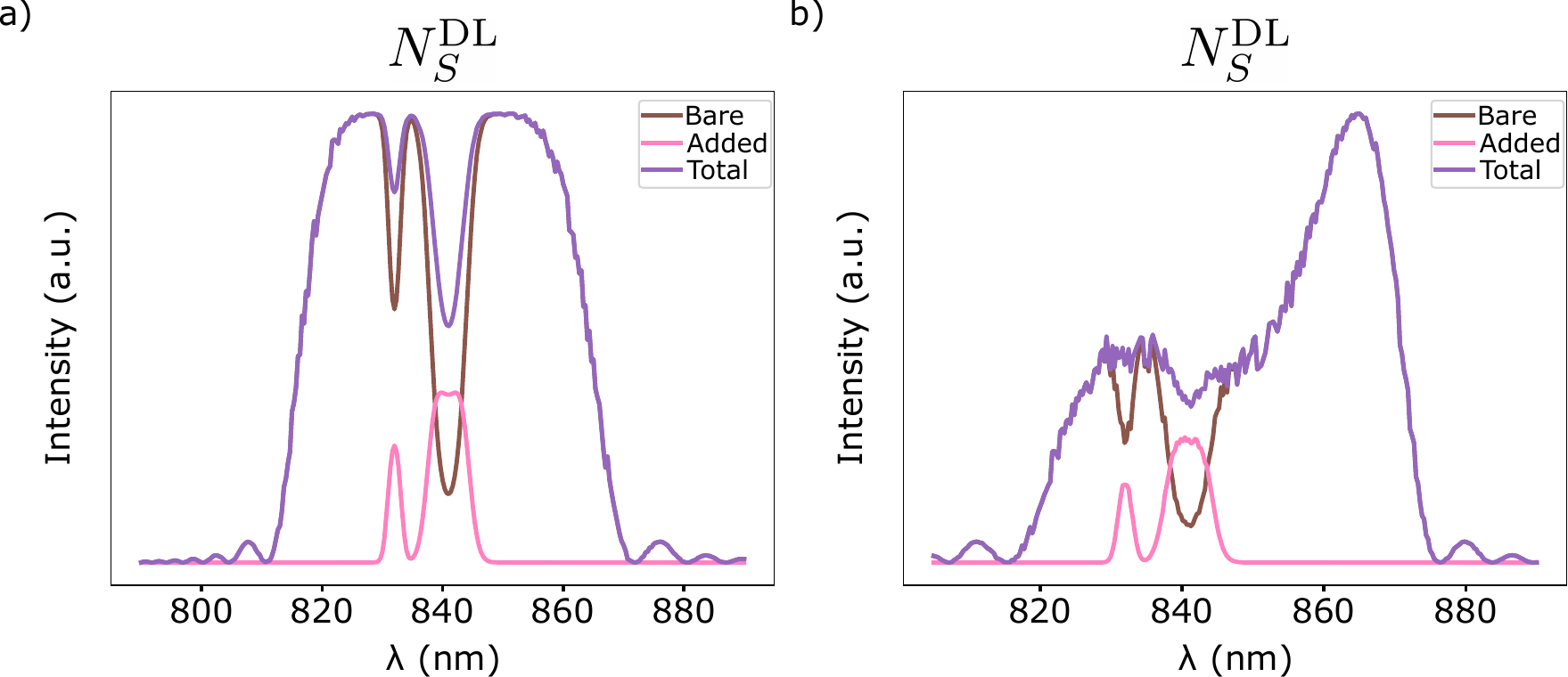}
		\caption{Distributed loss model signal intensity for the (a) flat parameters and (b) skewed parameters. We break the intensity down into it's different component: the ``bare'' and ``added noise'' contributions. The added noise contributions remain relatively similar for both sets of parameters, however, because of lack of phase-matching in the skewed parameter regime the loss effects on the bare contribution are less pronounced. Plots are shown for the low-gain regime.}
		\label{fig:contribution}
	\end{figure}

    In Figs.~\ref{fig:contribution}(a) and (b) we show the signal intensity and its two different contributions (bare and added noise) for the flat and skewed parameters, respectively. In both cases, we see that the relative effect of the added noise contribution is similar. In fact, the determining factor in the added noise contribution is the overall factor of $\kappa_{I}(-\omega)$ which multiplies the integral. On the other hand, when we consider the bare contribution, both parameter sets experience the same exponential suppression. However, since the skewed parameter set is not phase-matched in the plateau, the signal intensity's initial value is lower than the flat parameter set. This in turn leads to suppression valleys which are on the order of the added noise contribution for the skewed parameter and so when we add both contributions they almost cancel out. Note that the plots in Fig.~\ref{fig:contribution} are in the low-gain regime.  

     \section{Overall levels of idler loss}\label{sec:losslevel}
    
     The distributed model theory also gives rise to an interesting result concerning the overall levels of loss felt by the idler when we consider idler only loss. Although this is touched on in Ref.~\cite{houde2019loss}, we re-iterate the results here.

\begin{figure}[ht]
	 	     \includegraphics[width=1\linewidth]{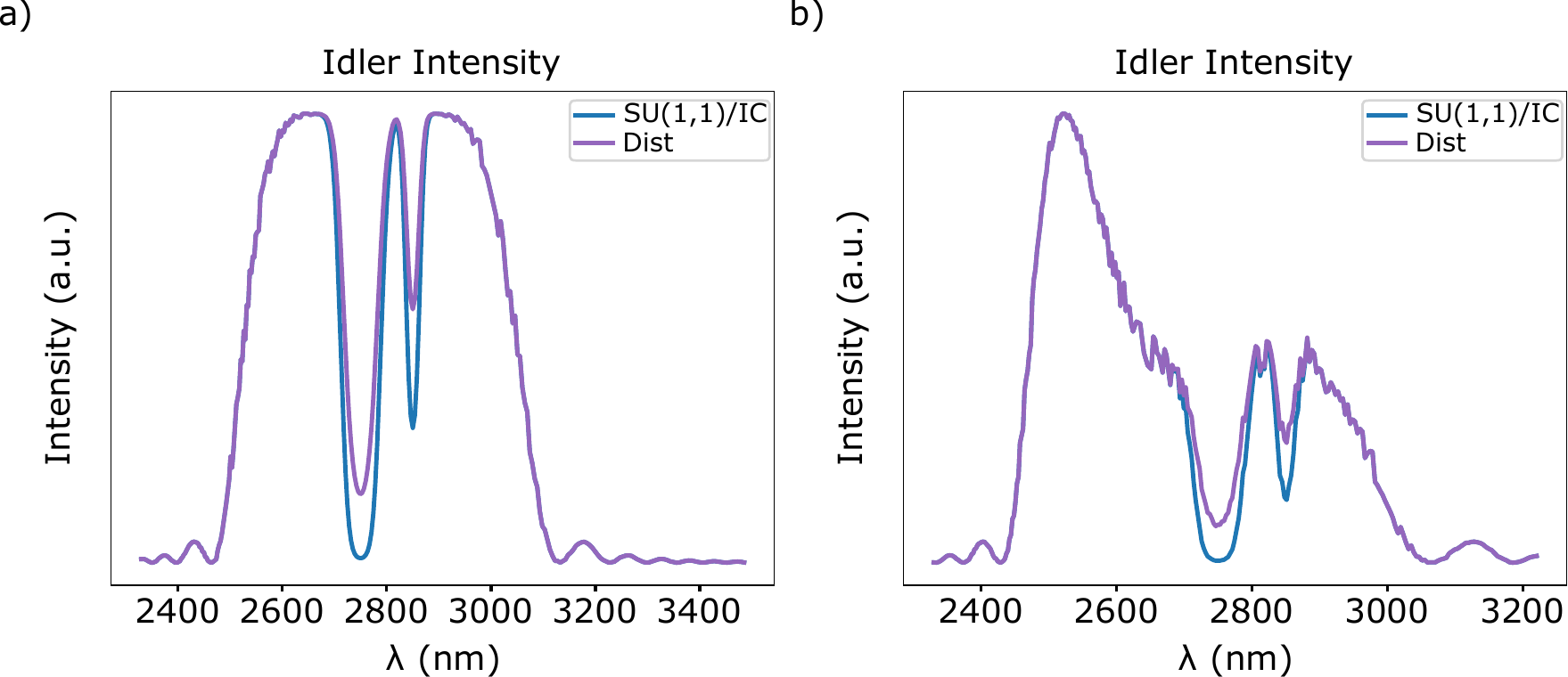}
	 	     \caption{Comparing the effects of idler loss on the idler intensity for all models for the (a) flat parameter set and (b) skewed parameter set. The idler loss is implemented as described in Sec.~\ref{sec:loss}. For the SU(1,1) and IC models, the intensity is evaluated after the beamsplitter loss but before the second pass. We see that in the distributed loss model, the idler experiences a lower level of overall loss due to the nonlinear interaction and mixing with the signal. Plots are shown for the low-gain regime.}
	 	\label{fig:loss_comparison}
	   \end{figure}

     When the nonlinear interaction is turned off, the distributed loss model is equivalent to a beamsplitter loss model where we relate the decay rates to transmission rates as done in Eq.~(\ref{eq:kappa_to_trans}) of the main text. Hence, one might naively expect the idler to experience the same levels of loss in both models (for the SU(1,1) model we consider the effects of loss after the beamsplitter but before the second pass). However, as we see in Fig.~\ref{fig:loss_comparison} this is not the case. In the distributed model, the overall levels of loss are lower than expected when the interaction is turned on (the idler intensity is higher than in the SU(1,1) model). This can be understood by the fact that since the loss is distributed, injected noise at the beginning of the nonlinear region will also be subject to the nonlinear interaction and as such will also be amplified/squeezed. This in turn leads to a higher level of idler photons at the end of the interaction. Furthermore, because the nonlinear interaction mixes the signal and idler modes as they propagate, the losses experienced by one mode also affect the other mode. 

     \section{SU(1,1) with OPD: skewed parameter set}\label{app:skewed_opd}
\begin{figure}[t]
		\includegraphics[width=1\linewidth]{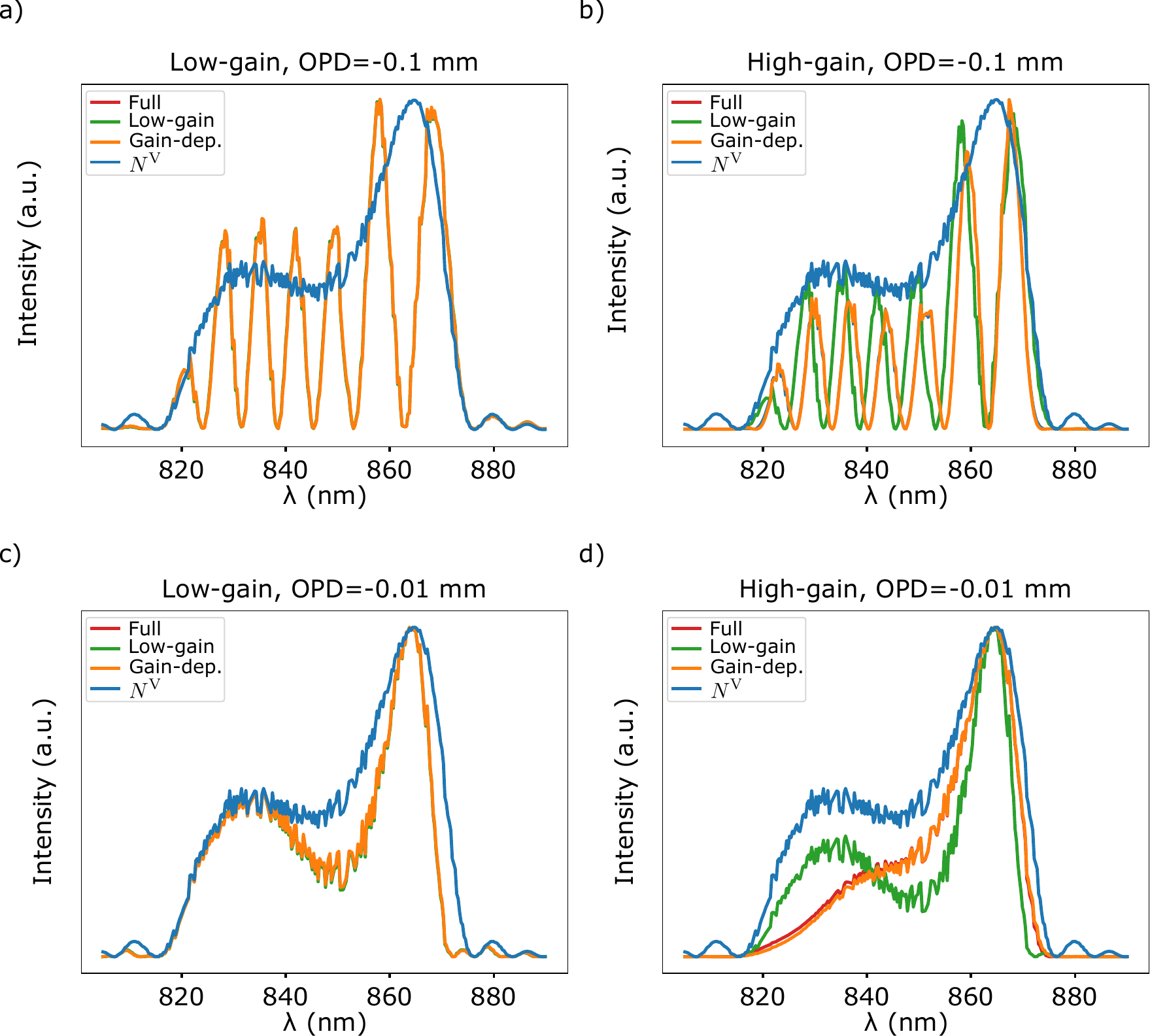}
		\caption{Normalized signal intensity for the SU(1,1) model with idler optical path delays (OPDs) and without any additional dispersion or loss for the skewed parameters. (a) and (b) represent the normalized signal intensities for the low- and high-gain regimes respectively, with an OPD of $-0.1$ mm. (c) and (d) represent the normalized signal intensities for the low- and high-gain regimes, respectively, with an OPD of $-0.01$ mm. The red curve is the full analytic solution (cf. Eq.~\ref{eq:ns_su11} of main text), the orange curve is the low-gain solution where the oscillations take the form of Eq.~\ref{eq:osc_low}, the green curve is the gain-dependent approximation described in Sec.~\ref{sec:su11_opd}, and the blue curve is the normalized signal intensity for a lossless single pass that we include to show the amplification region. The high-gain effects are more apparent for the skewed parameter set with an OPD of $-0.1$ mm.}
		\label{fig:opd_skewed}
	\end{figure}

    In this appendix, we consider the SU(1,1) interferometer where the idler experiences an OPD but no loss for the skewed parameter set. In Figs.~\ref{fig:opd_skewed}(a) and (b) we consider an idler OPD of $-0.1$ mm for the low- and high-gain regimes, respectively. As we can see from Fig.~\ref{fig:opd_skewed}(b) the high-gain effects are much more apparent for this set of parameters due to lack of phase-matching. In this regime, using the low-gain approximation to determine the additional idler dispersion would lead to erroneous results. Luckily, the gain-dependent approximation still follows the full solution quite well and can be used to infer the additional dispersion even for the skewed parameter set. In Figs.~\ref{fig:opd_skewed}(c) and (d) we show the low- and high-gain regimes, respectively, for an idler OPD of $-0.01$ mm. The results are similar to those of the flat parameter set included in the main text.

\bibliography{main}

\end{document}